\documentclass[aps,twocolumn,showpacs,preprintnumbers,prc,10pt,nofootinbib]{revtex4-1}

\usepackage{amsmath}
\usepackage{graphicx}

\begin{document}

\title{Isovector properties of the Gogny interaction}

\author{Roshan Sellahewa}
\affiliation{Department of Physics, Faculty of Engineering and Physical Sciences, University of Surrey, Guildford, Surrey GU2 7XH, United Kingdom}

\author{Arnau Rios}
\affiliation{Department of Physics, Faculty of Engineering and Physical Sciences, University of Surrey, Guildford, Surrey GU2 7XH, United Kingdom}

\date{\today}

\begin{abstract}
We analyse the properties of the Gogny interaction in homogeneous matter, with special emphasis on the isovector sector. We provide analytical expressions for both the single-particle and the bulk properties of symmetric and asymmetric nuclear matter. We perform an extensive analysis of these properties using 11 parametrizations extracted from the literature. 
We find that most Gogny interactions have low values for the slope of the symmetry energy, outside the range of empirically extracted values. As a test of extreme isospin dependence, we  also study the neutron star mass-radius relations implied by the different Gogny equations of state. Our results call for a more careful fitting procedure of the isovector properties of Gogny functionals.
\end{abstract}

\maketitle

\section{Introduction}

The Gogny force is a well-known and extensively used effective nuclear interaction \cite{Ring84}. Unlike the Skyrme density functional, which parametrizes the dependence on the relative distance  by contact interactions and derivatives, the Gogny force has a built-in finite range \cite{bender03}. This brings the Gogny force closer in spirit to realistic interactions. Moreover, the non-zero range is essential to avoid spurious truncations in the pairing channel within Hartree-Fock-Bogolyubov (HFB) nuclear structure calculations, and it was the main motivation behind its inception in the 1980's by the Bruy{\`e}res group \cite{Decharge1980}. In the decade that followed, Gogny forces were particularly used in nuclear fission studies \cite{Berger1991}. In this context, Gogny interactions are still a popular starting point for a variety of reasons \cite{Goutte2005}. Heavy nuclei, deformation and multipolar collective degrees of freedom have also been studied by using Gogny HFB \cite{Robledo2011,Gaffney2013,Peru2014}. Recently, even Gogny time-dependent calculations have become available \cite{Hashimoto2013}.

Here, we explore the predictions of the Gogny functional in a different context. Infinite nuclear matter has usually been taken as a reference in the fitting procedure of Gogny functionals \cite{bender03} (see below for a more detailed discussion). This includes isoscalar properties, such as saturation energies and saturation density. The compressibility of nuclear matter, of crucial importance for a variety of nuclear structure observables, has also been extensively studied with the Gogny functional \cite{Blaizot1995}. In contrast, the isovector properties of the Gogny parametrizations have hardly ever been discussed. Typically, only the symmetry energy is considered, if anything \cite{Decharge1980}. The remaining isovector dependence is expected to be captured by the fit to finite nuclei. Recent studies with the Skyrme functional, however, indicate that, in addition to nuclei, neutron-rich infinite matter is also needed to constrain the isovector sector \cite{Erler2013,Brown2014}. One could therefore put into question the predictive power of Gogny interactions in isovector-dominated properties, such as neutron skins, or neutron-rich systems. 

The isovector properties of Skyrme functionals have been extensively studied \cite{Stone2003,Steiner2005}, with a very wide variety of criteria to characterise their quality  \cite{Dutra2012}. Here, we aim at providing a generic description of isospin asymmetric nuclear matter with the Gogny interaction. Because only 11 functionals are available in the literature, one might think that this study is limited in the amount of variability, which is often referred to as \emph{systematic} uncertainty in the context of energy density functionals (EDFs) \cite{Dobaczewski2014,Goriely2014}. However, we find that, even within a relatively narrow set of Gogny functionals, there is a large variation in isospin properties. In particular, we observe that the density dependence of the symmetry energy provided by Gogny forces  is too soft and lies outside of currently accepted values \cite{Tsang2009,Tsang12}. This points to poor constraints in the isovector sector, which should be improved in future fitting protocols. 

We aim at finding general trends and, by providing analytical expressions, we hope to find specific combinations of parameters that might be responsible for critical behaviours. Where we can, we have compared with existing values of isoscalar and isovector properties \cite{Centelles2010,Chen2012}. As a specific aspect of the isovector sector, we  discuss neutron star properties as predicted by the present generation of Gogny forces. On the one hand, this might go beyond the scope of applicability of part of the Gogny functionals. On the other hand, a new generation of  observations is starting to put severe constraints on the equation of state of neutron-rich matter \cite{Steiner2010,Lattimer2012,Heinke2014}. Ideally, these constraints should also be considered in fitting procedures of energy density functionals \cite{Erler2013}. In line with the poor reproduction of bulk isovector properties, our calculations indicate that it is difficult to produce sufficiently massive neutron stars with the present generation of Gogny functionals. 

\section{Gogny interactions}

In terms of its functional form, the Gogny force is a natural extension of the early Brink and Boeker interaction \cite{Brink1967}. The finite-range part is modelled by two Gaussians, including a variety of spin-isospin exchange terms, as well as a zero-range density-dependent term that is helpful in reproducing saturation:
\begin{align}
V({\vec r})&=
\sum_{i=1,2} 
\Big( W_i+B_i\,P_{\sigma}-H_i\,P_{\tau}-M_i\,P_{\sigma}\,P_{\tau} \Big) e^{-\frac{r^2}{\mu_i}}  \nonumber \\
&+\sum_{i=1,2} t^i_0\,\left(1+x^i_0\,P_{\sigma}\right)\rho^{\alpha_i}\,\delta ({\vec r})  \nonumber
\\
&+ i W_0 (\sigma_1 + \sigma_2) [ {\vec k'} \times \delta({\vec r}) {\vec k}]
\, .
\label{eq:gogny_force}  
\end{align}
The first contribution includes the finite-range dependence, as $\vec{r}$ is the relative distance between two nucleons. All Gogny forces contain two terms (denoted by $i=1,2$) with effective ranges $\mu_1\approx 0.5-0.7$ fm and $\mu_2 \approx 1- 1.2$ fm, which in principle mimic a short- and a long-range component, respectively.\footnote{Note that these are generally fixed at the start of the fitting protocol and hence cannot be considered as fit parameters.} The spin-isospin structure of the force is relatively rich, and includes spin and  isospin exchange operators, $P_{\sigma}$  and $P_{\tau}$, respectively. The second contribution is a zero-range, density-dependent component that accounts for three-body correlations.\footnote{Note that, in most cases, the force contains a single $t_0$ term. The exception is the D1P parametrization, which includes two zero-range terms to make the fitting procedure more flexible \cite{Farine1999}.}
The force further incorporates a spin-orbit component,  proportional to $W_0$. This is a function of the relative momentum, $\vec{k}=(\nabla_1 - \nabla_2)/2i$, acting either on the bra or the ket ($\vec{k'}$) of two-nucleon states. Tensor terms in the Gogny functional have also been considered for a variety of applications \cite{Otsuka2006,Anguiano2012,Grasso2013}. Both terms depend on gradients of the density and are therefore irrelevant for nuclear matter bulk properties. We will not consider them hereafter.

There are about $14-17$ numerical parameters to be fit in a Gogny functional, although more often than not some of these are fixed at the outset of the fitting procedure.  In the following, we give results for the 11 Gogny parametrizations that we have been able to find in the literature. The original force, D1, was fit to the properties of closed-shell nuclei, $^{16}$O and $^{90}$Zr, as well as to nuclear matter saturation properties, including a relatively low saturation symmetry energy of $S=30.7$ MeV \cite{Decharge1980}. A new parametrization, D1S, was devised shortly after specifically for the study of fission \cite{Berger1991} and has been used extensively further \cite{Isayev2004,Goutte2005,Roca-Maza2011,Gaffney2013}. In an effort to pin down the bulk isoscalar properties of nuclear matter from nuclear data, including pairing correlations, Blaizot and collaborators formulated a series of Gogny interactions (D250, D260, D280 and D300) with a wide range of compressibilities \cite{Blaizot1995}. These have not been used extensively in the literature, but provide an interesting testing ground for isoscalar-dependent properties \cite{Rios2010a}.

Farine and collaborators conceived D1P as an extension of the usual Gogny functionals, increasing the number of zero-range terms to two~\cite{Farine1999}. Among other things, this extension improves the neutron matter equation of state by fits to realistic many-body calculations. Not surprisingly, we find that D1P performs well in the isovector sector. A similar idea is behind the D1N parametrization of Chappert et al.~\cite{Chappert2008,Chappert2007}, which also reduces the difference between theoretical and experimental masses in the actinide region. D1AS is an extension of D1, which has been used in the context of transport calculations \cite{Ono2003}. A major motivation for this force was to provide a stiffer symmetry energy, but its nuclear structure properties have not been explored to our knowledge. GT2, in contrast, was developed to provide realistic nuclear structure calculations including a tensor term, to account for changing shell structure in neutron-rich systems \cite{Otsuka2006}. Finally, D1M provides a global fit to masses of comparable quality to mass formulas within the HFB approach, including quadrupole correlation energies \cite{Goriely2009,Rodriguez2014}.

The finite range of the Gogny force is a more realistic feature, that is now customarily used in nuclear structure studies. In contrast, momentum dependence is incorporated into transport studies on a more intermittent basis \cite{Li2008}. In that context, one usually employs the so-called "momentum-dependent interactions" (MDI), which are vaguely related to Gogny forces. In particular, the momentum dependence and the isospin dependence are parametrized differently. In the following, we focus strictly on Gogny functionals, but some of the conclusions can be relevant for MDI-type interactions. 

For nuclear systems with different isospin contributions, like isospin-polarized matter, one can group the spin-isospin prefactors in the Gogny matrix elements into different terms. The zero-range contribution has a direct and an exchange part that, in practice, are computed together. For the finite-range terms, however, it is convenient to split the contribution into direct terms, which will be proportional to densities, and exchange terms, which involve more complicated functions of Fermi momenta. For the zero-range and direct terms, for instance, we differentiate between isoscalar ($0$ subscript):
\begin{align}
	A^i_0 &= \frac{\pi^{3/2} \mu_i^3}{4} \left[ 4 W_i +2 B_i - 2 H_i - M_i \right] \, , \label{eq:as} \\
	C^i_0 &= \frac{3}{4} t_0 \, ,
\end{align}
and isovector ($1$ subscript):
\begin{align}
	A^i_1 &= \frac{\pi^{3/2} \mu_i^3}{4} \left[ -2 H_i - M_i  \right] \, , \\
	C^i_1 &= - \frac{1}{4} t_0 \left( 1+2 x_0 \right) \, ,
\end{align}
contributions. For the finite-range exchange contribution, it is useful to consider terms associated with equal and unequal isospin pairs:
\begin{align}
	B^i_{nn} = B^i_{pp} &= - \frac{1}{ \sqrt{\pi} } \left[  W_i +2 B_i - H_i - 2 M_i \right] \, , \\
	B^i_{np} &=  \frac{1}{\sqrt{\pi}} \left[   H_i + 2 M_i \right] \, . \label{eq:bnp}
\end{align}
In discussing isoscalar and isovector single-particle properties, we also introduce the exchange isoscalar and isovector terms:
\begin{align}
	B^i_0 &= 	B^i_{nn} + B^i_{np} \nonumber \\
	& =- \frac{1}{ \sqrt{\pi} } \left[  W_i +2 B_i - 2H_i - 4 M_i \right] \, , \\
	B^i_1 &= 	B^i_{nn} - B^i_{np} = - \frac{1}{ \sqrt{\pi} } \left[  W_i +2 B_i \right] \, . 
\end{align}
In infinite matter, and within the Hartree-Fock approximation employed in this work, the single-particle (bulk) properties have contributions associated with the exchange term which are proportional to single (double) integrals of gaussians over the Fermi surfaces of neutrons and protons.  These integrals can be computed analytically and give rise to a series of polynomial and gaussian functions. To avoid cluttering our discussion with equations, we provide all analytical expressions in the Appendixes. 

\begin{figure}
\begin{center}
\includegraphics[width=0.9\linewidth]{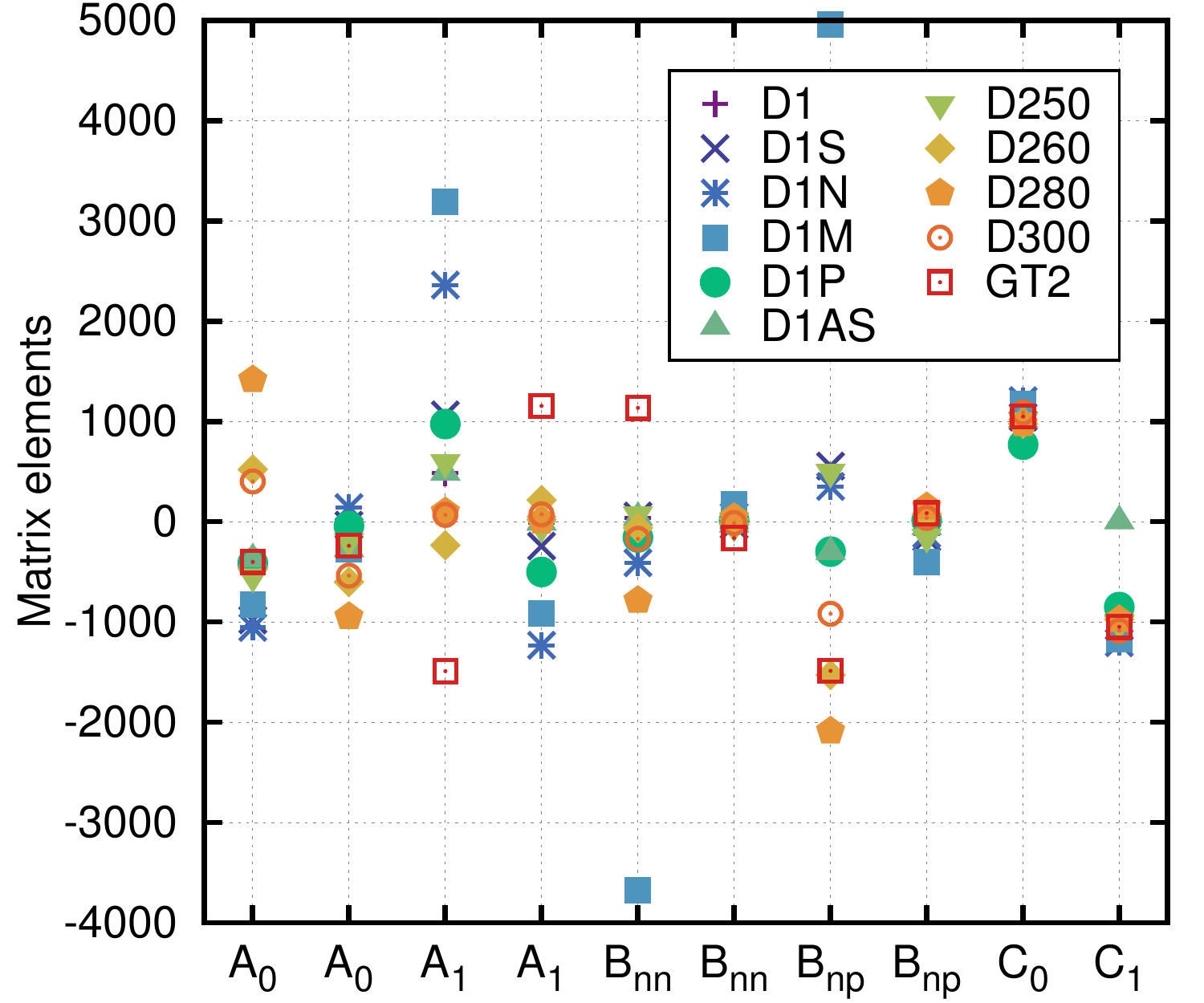}
\caption{(Color online) Matrix elements of Eqs.~(\ref{eq:as}) to (\ref{eq:bnp}) for the 11 Gogny functionals under consideration. The first (second) coefficient of every couple corresponds to $i=1$ ($i=2$). For the zero range terms, $C_0$ and $C_1$, we only present $i=1$. The units are MeVfm$^{3}$ for $A_i$'s;  MeV for $B_i$'s and MeVfm$^{3\alpha_i+3}$ for $C_i$'s.}
\label{fig:matrix_elements}
\end{center}
\end{figure}

The numerical parameters appearing in Eq.~(\ref{eq:gogny_force}) are obtained by a fitting procedure of the Gogny functional. With these, one can compute the different matrix elements. We note that the parameters can be degenerate, in the sense that only linear combinations enter the fitting procedure. In addition, the separation of matrix elements in zero-range, direct finite-range and exchange terms is arbitrary. Hence, the independent values of these matrix elements are not necessarily meaningful. Some specific parameters, however, do determine physical properties (see below for how $B^{nn}$ is entirely responsible for the effective mass splitting), and hence it can be interesting to find out their values. While a detailed analysis is beyond the scope of this work, we provide a plot with the values of these matrix elements in Fig.~\ref{fig:matrix_elements}. This provides, at a glance, an explanation of the different isovector and isoscalar parameters for the 11 functionals under consideration. 

There are, for instance, common trends that can impact isoscalar and isovector properties of matter. Most forces prefer a large positive $A^1_1$ and a negative $A^2_1$, suggesting cancellations in the isovector, direct finite-range part of the functional. In contrast, the majority of forces prefer negative  $B_{xy}^{1,2}$, which suggests that the exchange terms act as overall attractive  contributions. The isospin singlet zero-range term $C_0$ is repulsive, as expected from the usual density-dependent terms of the functionals. In contrast, all forces, except for D1AS, present attractive $C_1$ contributions. This suggests a dominance of attractive terms in the isovector channels which, as we shall see, hampers the development of stiff symmetry energies. 

In addition, we find a relatively large spread for most parameters. This is a sign of large functional dependence or systematic uncertainty \cite{Dobaczewski2014}. In particular, all the short-range parameters $A^1_x$ and $B^1_x$ present a much larger variability than their long-range counterparts, $A^2_x$ and $B^2_x$. In terms of functionals, D1M is an outlier as compared to most other parametrizations, with extreme values of finite-range exchange parameters, $B^1_{nn}$ and $B^1_{np}$. The specific optimisation procedure of this force should most likely account for these large differences \cite{Goriely2009}. Similarly, GT2 also shows a distinct behaviour for $A^{1,2}_1$ and $B^{1,2}_{nn}$, as already acknowledged in the original publication \cite{Otsuka2006}. The isoscalar zero range matrix elements $C^1_0$ are, as expected, all repulsive, whereas their isovector counterparts, $C^1_1$, are attractive and of a similar order of magnitude.\footnote{All forces have $\alpha_i=\tfrac{1}{3}$ except for D250 and D300 which have $\alpha_i=\tfrac{2}{3}$.} We do not show the $C^2_x$ parameters, since they are zero for all forces except for D1P, where they are repulsive $C^x_2 \approx 192$ MeV fm$^{4}$. 

\section{Microscopic properties}

\subsection{Single-particle potentials}

We start our discussion by looking at a series of single-particle properties of asymmetric nuclear matter as obtained by different Gogny functionals at different densities. All these properties characterise, in one way or another, the single-particle potential of a neutron or a proton with momentum $k$, denoted by $U_\tau(k)$. The isospin index, $\tau$, corresponds to a neutron, $\tau=+$, or a proton, $\tau=-$. We work within the Hartree-Fock approximation and, in asymmetric infinite matter, the single-particle potential is the result of an integral and spin average over the neutron and proton Fermi surfaces:
\begin{align}
U_\tau(k) &= \frac{1}{2}\sum_{\substack{\sigma',\tau' \\ \sigma}}  \int \frac{\text{d}^3 k'}{(2 \pi)^3}
\left\langle k \sigma \tau; k' \sigma' \tau' \right| V \left| k \sigma \tau; k' \sigma' \tau' \right\rangle_A \, ,
\end{align}
so that the integral runs only for $k'<k_F^{\tau'}$. The subscript $A$ denotes an antisymmetrization in the matrix element. 

Using the expression for the Gogny functional in Eq.~(\ref{eq:gogny_force}), one can find analytical expressions for $U_\tau(k)$. The zero-range and the direct terms, for instance, are momentum independent and can be integrated straight away. This gives rise to two terms proportional to the density, or to $\rho^{\alpha_i+1}$, in the case of the zero-range contribution. The exchange term, arising from the antisymmetrization, is more cumbersome. It involves an integral over a gaussian momentum factor, which includes at least one angular integration. This integral can be computed in a closed form, and we provide the explicit expression in  the Appendix [Eq.~(\ref{eq:sp_potential_app})].
The isovector contribution is proportional to the isospin-asymmetry fraction $\beta=\frac{\rho_n-\rho_p}{\rho}$. The result of the integral in the exchange finite-range contribution depends on the Fermi momenta of each species, $k_F^\tau = \left( \frac{3 \pi^2}{2} \rho \left[ 1 \pm \beta \right] \right)^{1/3}$. We do not provide the rearrangement term, Eq.~(\ref{eq:rear}), explicitly here, but we include it in all figures that need it. 
The function $\mathsf{u} \left( q, q_F \right)$, given in Eq.~(\ref{eq:ufun}), involves gaussians and error functions. This encodes the momentum dependence of the single-particle potential, as all the other terms are constants as a function of $k$. 

\begin{figure}[t!]
\begin{center}
\includegraphics[width=0.9\linewidth]{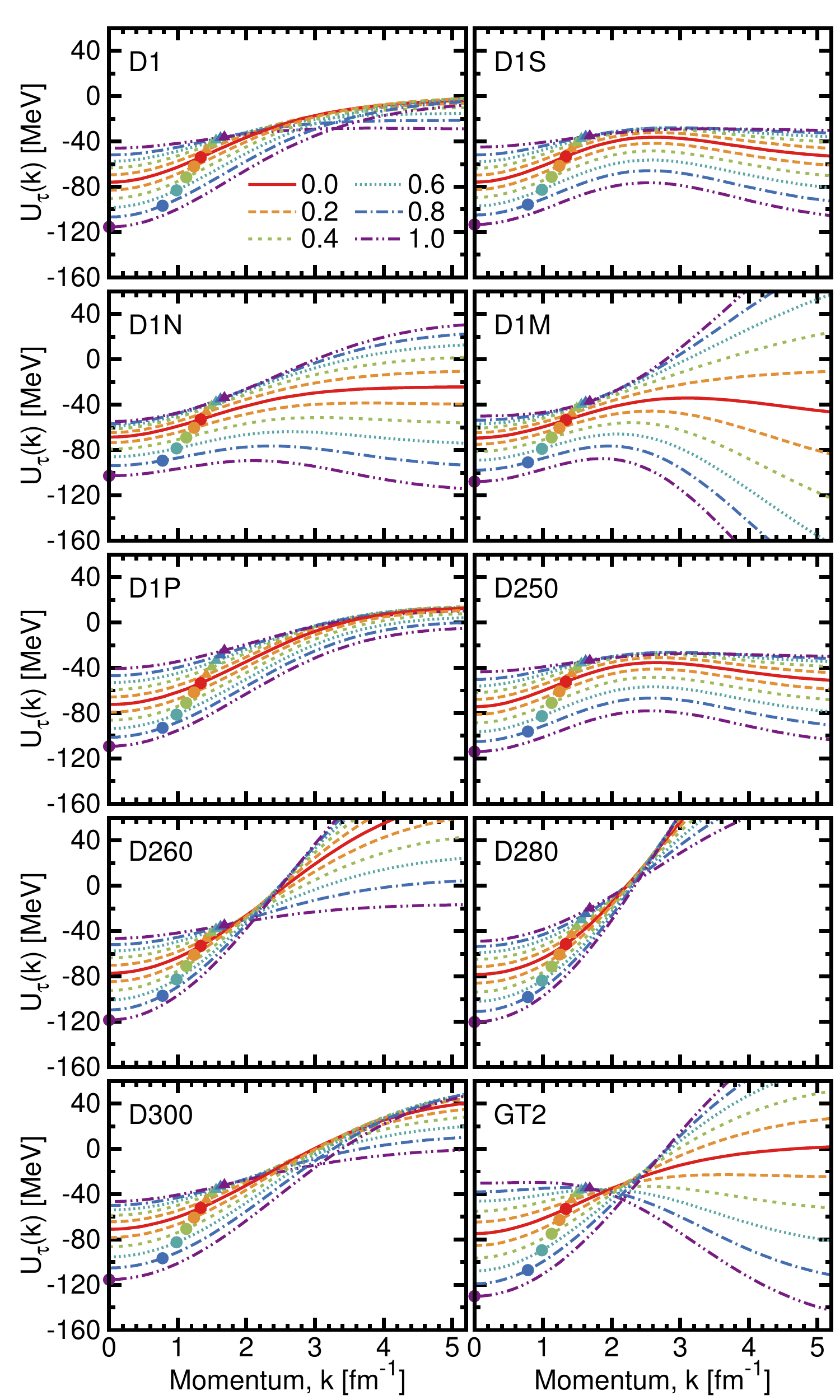}
\caption{(Color online) Single-particle potential of neutrons (solid triangles) and protons (solid circles) as a function of momentum for 6 different isospin asymmetries: $\beta=0$ (solid line), $0.2$ (long dashed line), $0.4$ (short dashed line), $0.6$ (dotted line), $0.8$ (dashed-dotted line) and $1.0$ (double-dotted dashed line). The results were obtained at $\rho=0.16$ fm$^{-3}$. The symbols denote the single-particle potential of a neutron (triangles) or a proton (circles) at the respective Fermi momentum, $k_F^\tau$. }
\label{fig:uk_tau}
\end{center}
\end{figure}

We show the single-particle potentials, $U_\tau(k)$, as a function of momentum in Fig.~\ref{fig:uk_tau}. The different panels correspond to 10 different Gogny parametrizations.\footnote{We ignore D1AS for the time being, because its momentum dependence is identical to D1.} The results have all been computed at $\rho=0.16$ fm$^{-3}$ at different isospin asymmetries (see Figure caption for details). We highlight with symbols the neutron (triangles) and proton (circles) potentials at the respective Fermi surfaces. These contributions are relevant for our understanding of the evolution of isospin in single-particle properties.

The single-particle potential for symmetric matter (solid lines) is rather well constrained at low momenta. All forces predict values $U_\tau(0) \approx 70-80$ MeV at zero momentum and $\rho=0.16$ fm$^{-3}$. At low momentum, below about $2$ fm$^{-1}$, the symmetric matter single-particle potentials are similar. As a function of momentum, $U(k)$ generally increases with $k$ in this region. Above $k\approx2$ fm$^{-1}$, for some functionals (D1N, D1P or GT2) the potential saturates, decreases (D1S, D1M, D250) or increases (D260, D280, D300). For symmetric matter, the momentum dependence is dictated by the sum of two terms, governed by $B_0^i$ and the function $\mathsf{u} \left( q, q_F \right)$ evaluated at the Fermi momentum, $q_F= \mu_i k_F$, of symmetric matter. We note that in the $k \gg 1$ limit, the momentum-dependent exchange term becomes negligible and the direct plus zero-range terms dominate.
Hence, the high-momentum value is entirely dominated by momentum-independent terms.

The interplay between asymmetry and momentum is also relevant, particularly in transport calculations \cite{Chen2012,Li2013,Coupland2014}. The data presented in Fig.~\ref{fig:uk_tau} shows that, at low momentum, the isospin asymmetry dependence of single-particle potentials is relatively well constrained, at least around saturation density. In all cases, we observe an increase of the neutron $U_n(k)$ as a function of asymmetry, whereas the proton $U_p(k)$ decreases. This corresponds to the physically intuitive idea that neutrons (protons) are less (more) bound in neutron-rich systems. In the low momentum region, below $\approx 2$ fm$^{-1}$, the dependence in asymmetry is rather monotonic. For an increase of $0.2$ in asymmetry, we find a steady decrease of $U_p$ by around $\approx 10$ MeV. In contrast, the neutron potential increases by about $5$ to $7$ MeV in a pattern that is less linear. This suggests that, in the limit of neutron-rich systems, the non-linear exchange terms dominate the isospin dependence of the single-particle momentum. For neutrons in neutron matter, the single-particle potential at the Fermi surface is $U_n \approx -32$ MeV, with a spread of around $5$ to $10$ MeV. In the limit of extreme isospin imbalance ($\beta=1$), we find that most forces predict a similar value for the zero-momentum proton potential, $U_p(0) \approx -115$ MeV, with a spread of around $10$ MeV.  A proton impurity is a particularly interesting system, because its momentum dependence is entirely governed by $B^i_{np}$ \cite{Forbes2014,Roggero2014}.

The asymmetry dependence in the high momentum region, $k \gtrsim 2$ fm$^{-1}$, is less constrained. One finds a large variety of results. For D1, for instance, the single-particle energy of neutrons for $k>3$ fm$^{-1}$ is lower than that of protons. This inversion occurs also for D260, D270, D300 and GT2 in a region ranging between $2$ and $3$  fm$^{-1}$. The results obtained with D1S, D1P and D250 suggest that the asymmetry dependence of the neutron potential is very weak above $2.5$ fm$^{-1}$. In stark contrast, D1M suggests a strong increase (decrease) of $U_n$ ($U_p$) with asymmetry in the large momentum region. As a matter of fact, for this force at large asymmetries, $U_p$ decreases rather steeply as a function of momentum. We will see the consequences of this behaviour in the analysis of the effective mass that follows. All in all, this figure suggests that the isospin asymmetry dependence of the high-momentum single-particle properties is not constrained in the Gogny functional. One could foresee improvements in this direction by using fitting protocols that took into account the information available from realistic many-body calculations in isospin asymmetric nuclear matter \cite{Zuo2002,Frick2005,Vidana2009}.

\begin{figure*}[t]
\begin{center}
\includegraphics[width=0.7\linewidth]{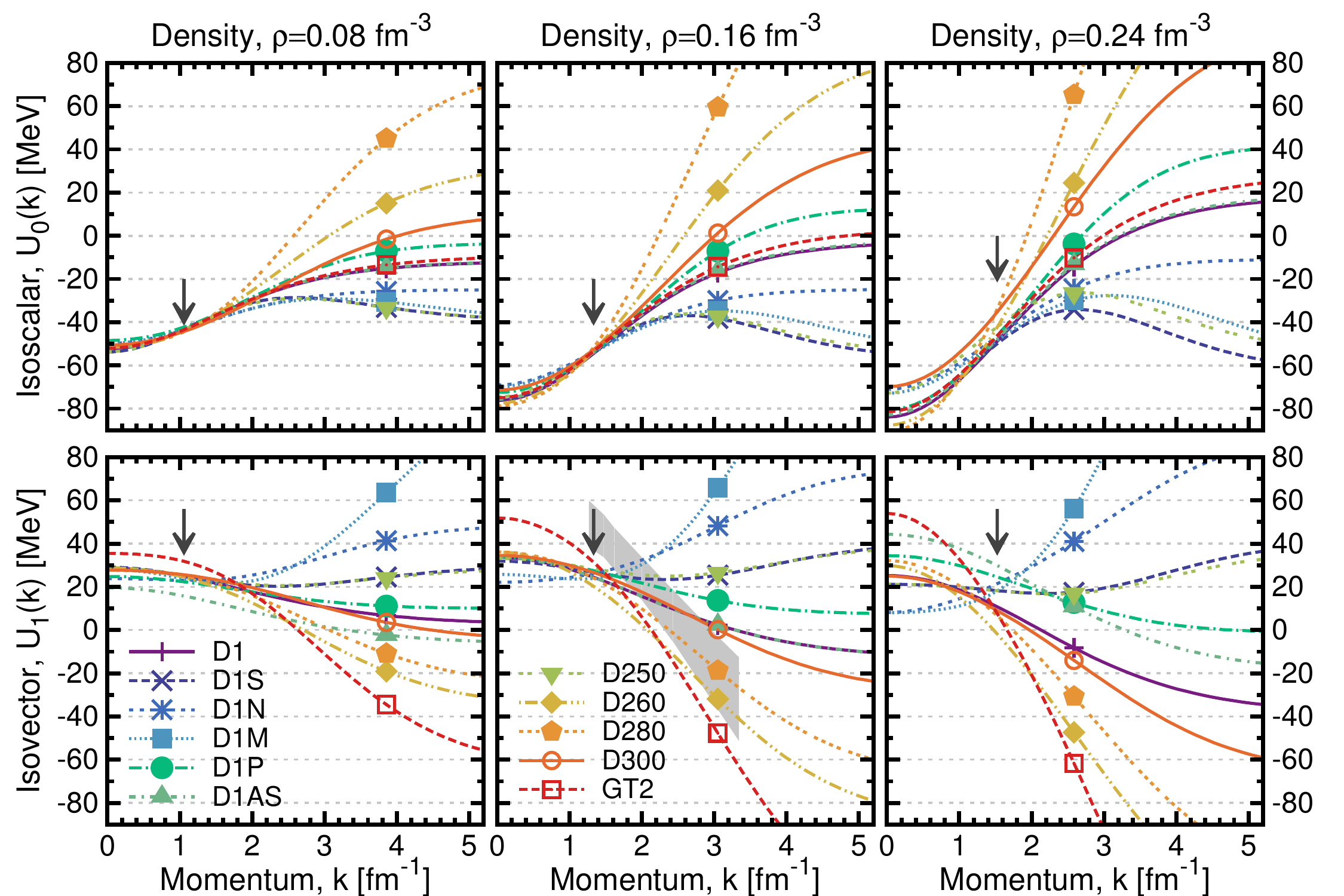}
\caption{(Color online) Isoscalar (top panels) and isovector (bottom panels) components of the single-particle potential as a function of momentum. Results for all Gogny functionals are displayed at 3 densities: $\rho=0.08$ fm$^{-3}$ (left panels), $\rho=0.16$ fm$^{-3}$ (central panels) and $\rho=0.24$ fm$^{-3}$ (right panels) are displayed. The gray band in the bottom central panel is the allowed region of saturation isovector single-particle potentials obtained in Ref.~\cite{Li2014}. The arrows mark the position of the Fermi momentum at each density.}
\label{fig:u_iso_sc_vec}
\end{center}
\end{figure*}

Up to this point, we have only displayed results computed at a single density, $\rho=0.16$ fm$^{-3}$. One would like to know whether similar issues are found at different densities. Rather than showing the whole asymmetry dependence of the single-particle potentials for different densities, we opt for displaying separately the isovector and isoscalar components of the single-particle potential. As described in Appendix~\ref{app:micro}, one can introduce an isoscalar potential, $U_0(k)$, which is essentially the average of the neutron and proton potentials. This basically corresponds to the single-particle potential of symmetric matter, which one would expect to be well constrained at sub-saturation densities by nuclear data. The isovector component, $U_1^\text{sym}(k)$ [see Eq.~(\ref{eq:u_1_sym}) for a definition], is the first derivative with respect to $\beta$ of the single-particle potentials. It therefore encodes information on how the potentials evolve with asymmetry. We note that this isovector potential is essentially equivalent to the Lane potential in a wide density, asymmetry and momentum range \cite{Charity2014,*Charity2014a,Li2014}. 

We show the isoscalar potential for three characteristic densities in the top panels of Figure~\ref{fig:u_iso_sc_vec}. At half saturation (top left panel), the low-momentum part of the single-particle potential is very well constrained, in the sense that all functionals predict very similar values. At $k=0$, for instance, one finds $U_0(0) \approx -50$ MeV. The increase of $U_0(k)$ with momentum is rather mild and, at $k=2$ fm$^{-1}$, most potentials are around $-30$ MeV. As momentum increases past this point, though, a relatively large spread of values develops. 

At saturation (central panels), one finds qualitatively equivalent results. The low-momentum region of the potential is well constrained, although larger differences between functionals are observed. In the region $k \approx 1$ fm$^{-1}$, all isoscalar potentials are close to $U_0 \approx -60$ MeV. Above this momentum region, large divergences appear. Whereas some functionals increase indefinitely with momentum, others saturate or even decrease. For densities above saturation, in contrast, there is a relatively wide spread of potentials both at low and at high momenta. For instance, the zero-momentum single-particle potential is predicted to range between $-90$ and $-70$ MeV. As with other densities, the spread increases substantially above $k>2$ fm$^{-1}$. 

Regarding the momentum dependence of the different functionals, it is interesting to note that it is (up to density-dependent normalizations) essentially the same at different densities. D1S, for instance, predicts at all density an isoscalar potential that increases, then saturates and further on decreases with momentum. In contrast, the D1N isoscalar potential increases monotonicly at low $k$ and saturates rather quickly at all densities. D280 is extreme, in that its momentum dependence is the steepest, with a large increase in $U$ as $k$ becomes larger. 

We note that the functionals with largest $U_0$ at high momentum (D280, D260 and D300) are those that displayed large $A_0^1$ values in Fig.~\ref{fig:matrix_elements}. At large $k$, one expects the exchange term in $U_0$ to become negligible. The isoscalar single-particle potential hence tends to the momentum-independent value:
\begin{align}
	U_0 (k \gg 1) \approx \sum_{i=1,2} \left[ A^i_0 + C^i_0 \rho^{\alpha_i} \right] \rho \, .
\end{align}
Forces with large $A^x_0$ and $C^x_0$ matrix elements will eventually show large momentum-independent contributions in the high-momentum region of $U_0(k)$ Moreover, since most zero-range isoscalar couplings $C^1_0$ are positive, those functionals that have large and positive $A^1_0$ will develop strongly repulsive single-particle potentials at high momenta and large densities. Note that D1P was precisely fit to have $U_0(k)$ cross zero around $k \approx 3.2$ fm$^{-1}$ and tend to an asymptotic value of $30$ MeV for $k \gg 1$ \cite{Farine1999}. 

The bottom panels of Fig.~\ref{fig:u_iso_sc_vec} show a strikingly different pattern.  Even at sub-saturation densities (bottom left panel), the low-momentum components of the isovector potentials coming from different functionals are rather different. At zero momentum, we find values ranging from $U_1(0) \approx 19$ to $35$ MeV, with significant divergences. At saturation (bottom central panel), most functionals predict $U_1(0) \approx 33$ MeV, except for GT2, D1M and D1N. Between $k=1.4$ and $1.6$ fm$^{-1}$, there is an area of overall agreement between functionals, with values of $U_1 \approx 25$ MeV, but the results diverge even more than in the isoscalar sector as momentum increases. The isovector potentials above saturation (bottom right panel) cover a wide range of values, which indicates that they are not constrained by the parameter fitting procedure. 

A recent analysis of the isovector optical potential and its connection to bulk properties suggests that $U_1$ should decrease with momentum \cite{Li2013}. We show with a grey band the allowed region of the saturation density isovector single-particle potential as obtained from the optical potential fits of Ref.~\cite{Li2014}. At low momentum, the allowed region is above all single-particle potentials. The steep decrease as a function of momentum suggested by this analysis is only reproduced by a minority (D260, D280 and GT2) of extreme functionals. 
Empirical values and theoretical predictions seem to have somewhat similar momentum dependences, but the absolute values of $U_0(k)$ are somewhat too low. This shift in absolute values could be due to the lack of non-locality corrections in $U_0(k)$ \cite{Charity2014,*Charity2014a,Li2014}.  We note that our results for $U_0$ and $U_1$ agree with those presented in Ref.~\cite{Chen2012} for the corresponding parametrizations.

The high momentum asymmetry of $U_1$ is also determined entirely by the direct and zero-range matrix elements:
\begin{align}
	U_1 (k \gg 1) \approx \sum_{i=1,2} \left[ A^i_1 + C^i_1 \rho^{\alpha_i} \right] \rho \, .
\end{align}
Hence, the extremely large and positive $A^1_1$ values of D1M and D1N dictate limiting values of $U_1$ which are large, positive and increasing with density. In contrast, because $A^1_1 \ll 0$ for GT2, its isovector potential becomes very negative as momentum increases. 

To some extent, the large differences among functionals are not surprising. The fitting procedure includes only a series of points of finite nuclei, which are typically sub-saturation systems, and bulk, zero-temperature saturation matter properties. All of these data are essentially determined by single-particle properties at (a) densities below saturations and (b) momenta below the Fermi momentum, which is typically of the order $k_F \approx 1-1.3$ fm$^{-1}$. In addition, since most of these systems are almost isospin-symmetric, the single-particle isovector properties are rather poorly constrained. Consequently, the single-particle potential is only well constrained at low momentum, below saturation and near symmetric systems. A covariance analysis that propagates the statistical uncertainties of the fitting procedure would further quantify this statement \cite{Dobaczewski2014}.

\subsection{Effective masses}

The effective mass provides a sensitive characterisation of the momentum dependence of the single-particle potential. We note again that the momentum dependence of the mean-field is exclusively due to the exchange term, as both the direct and zero-range contributions are constants as a function of $k$. In other words, the function $\mathsf{u} \left( q, q_F \right)$, given in Eq.~(\ref{eq:ufun}) is entirely responsible for the non-trivial effective mass. As a consequence, the effective mass, $m_\tau^*$, is only proportional to the matrix elements $B^{nn}$ and $B^{np}$:
\begin{widetext}
\begin{align}
\frac{m_N}{m_\tau^*} =  1 + \frac{m_N}{\hbar^2 k} \frac{\partial U_\tau(k)}{\partial k} 
= 1 + \frac{m_N}{\hbar^2} \sum_{i=1,2} \Big\{ B^i_{nn} \mathsf{m} \left( \mu_i k, \mu_i k_F^{\tau} \right) + 
B^i_{np} \mathsf{m} \left( \mu_i k, \mu_i k_F^{-\tau} \right)  \Big\} \, . 
\label{eq:eff_mass}
\end{align}
\end{widetext}
The function $\mathsf{m} \left( q, q_F \right)$, given in Eq.~(\ref{eq:mfun}), is essentially a momentum derivative of the $\mathsf{u} \left( q, q_F \right)$ appearing in the single-particle potential.

\begin{figure}[t]
\begin{center}
\includegraphics[width=0.9\linewidth]{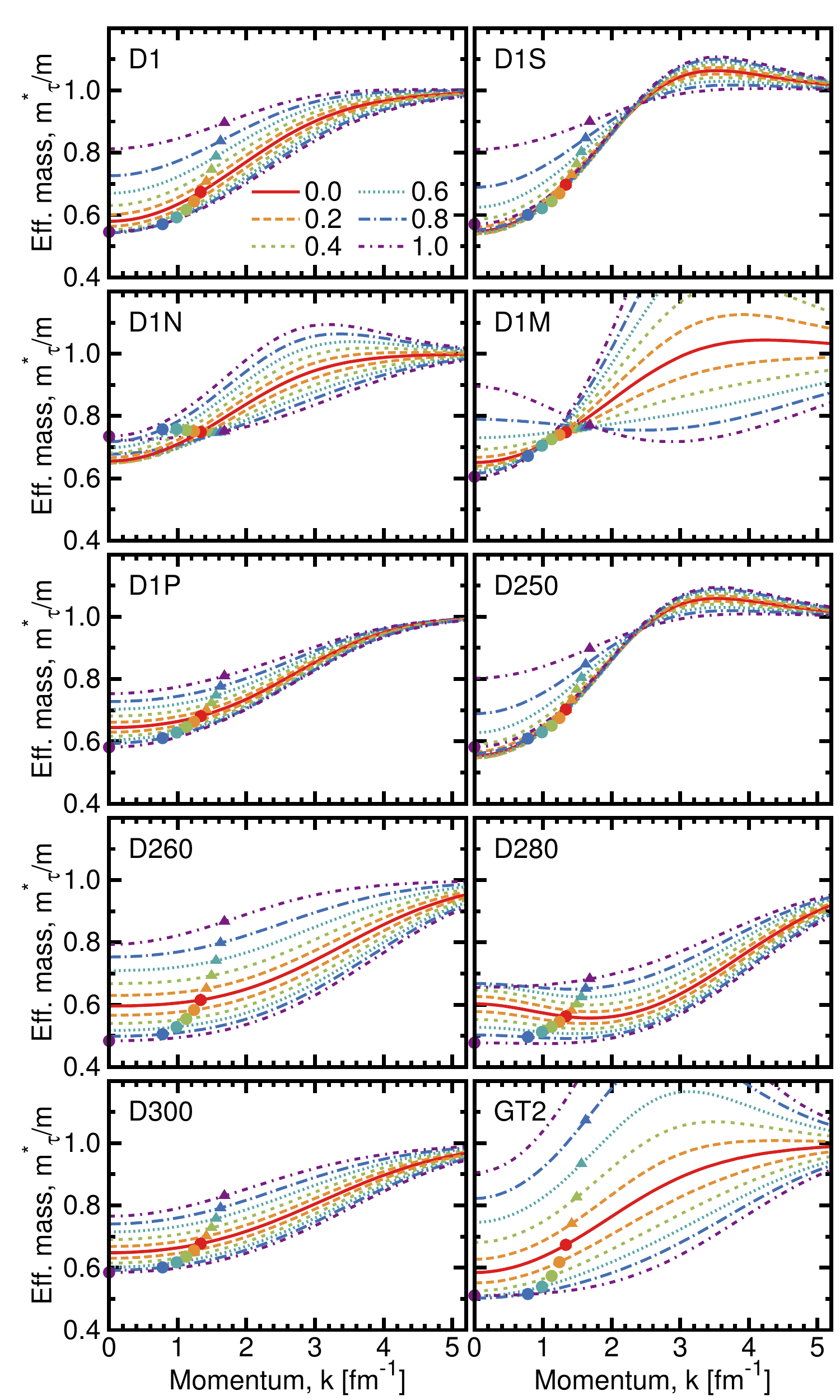}
\caption{(Color online) Effective mass of neutrons (solid circles) and protons (solid triangles) as a function of momentum for 6 different isospin asymmetries. The results were obtained at $\rho=0.16$ fm$^{-3}$.  See Fig.~\ref{fig:uk_tau} for further explanations.}
\label{fig:effm_tau}
\end{center}
\end{figure}

In analogy to Fig.~\ref{fig:uk_tau}, we present in the 10 panels of Figure \ref{fig:effm_tau} the results for the effective mass as a function of momentum at $\rho=0.16$ fm$^{-3}$ for different functionals. We explore different asymmetries and show the corresponding Fermi momentum effective masses of neutrons (protons) with solid triangles (solid circles). D1AS is a momentum independent extension of D1, so their effective masses are the same.

For isospin symmetric matter, $\beta=0$, and the corresponding saturation density, we provide the values of the effective mass of the 11 functionals in column 1 of Table~\ref{table:nm_properties}. In this case, the results of Fig.~\ref{fig:effm_tau} indicate that all effective masses at the Fermi surface are similar. We note, however, that the momentum dependence of $m^*_\tau(k)$ is relatively disparate for the different functionals. D1 shows a steady increase, saturating to $m^* \approx m$ at large momenta. Qualitatively similar results are found for D1N, D1P, D260, D300 and GT2. In contrast, D1S, D1M and D250 go through a maximum at $k \approx 3.5$ fm$^{-1}$ before settling down at $m^* \approx m$. D280 is unique, in that the effective mass at low momentum is a decreasing function of momentum.

The asymmetry dependence of the effective masses is also rather heterogeneous. For most forces, the neutron effective mass increases with asymmetry at low momentum. For D1P, for instance, an increase of $0.2$ in asymmetry results in an increase in the effective mass of around $0.02$. The decrease in proton effective mass is slightly smaller.  If we exclude D1N, the effective mass of a proton impurity in neutron matter at $k=0$ falls within $m^*_p(k=0) \approx 0.48$ to $0.62$. In contrast, the neutron effective mass in neutron matter changes from $m^*_n(k_F^n) \approx 0.68$ to $1.26$. Again, we point out that the impurity system could be used to constrain the value of $B^i_{np}$ \cite{Forbes2014,Roggero2014}.

D1N and D250 are exceptional, in the sense that their neutron effective masses at low momenta hardly depend on asymmetry. The momentum dependence of the neutron effective mass for D1M is striking, because it changes drastically from low to large asymmetries. In particular, the proton effective mass at large momenta increases substantially, to values above $m_p^* \approx 1.7 m_N$ in neutron matter. GT2 (bottom right panel) shows a similarly large dependence on asymmetry, but in this case it is the neutron effective mass that grows substantially with asymmetry. 

Note also that D1S, D1M and D250 show a crossing point, above which the neutron effective mass is smaller than the proton effective mass. Again, because this high-momentum, high-asymmetry region is relatively unconstrained, it is not surprising to find this variety of results. It would be interesting to explore whether this reordering with momentum has any impact on transport calculations \cite{Li2008} and observables of low energy nuclear collisions \cite{Coupland2014}.

\begin{figure}[t]
\begin{center}
\includegraphics[width=0.7\linewidth]{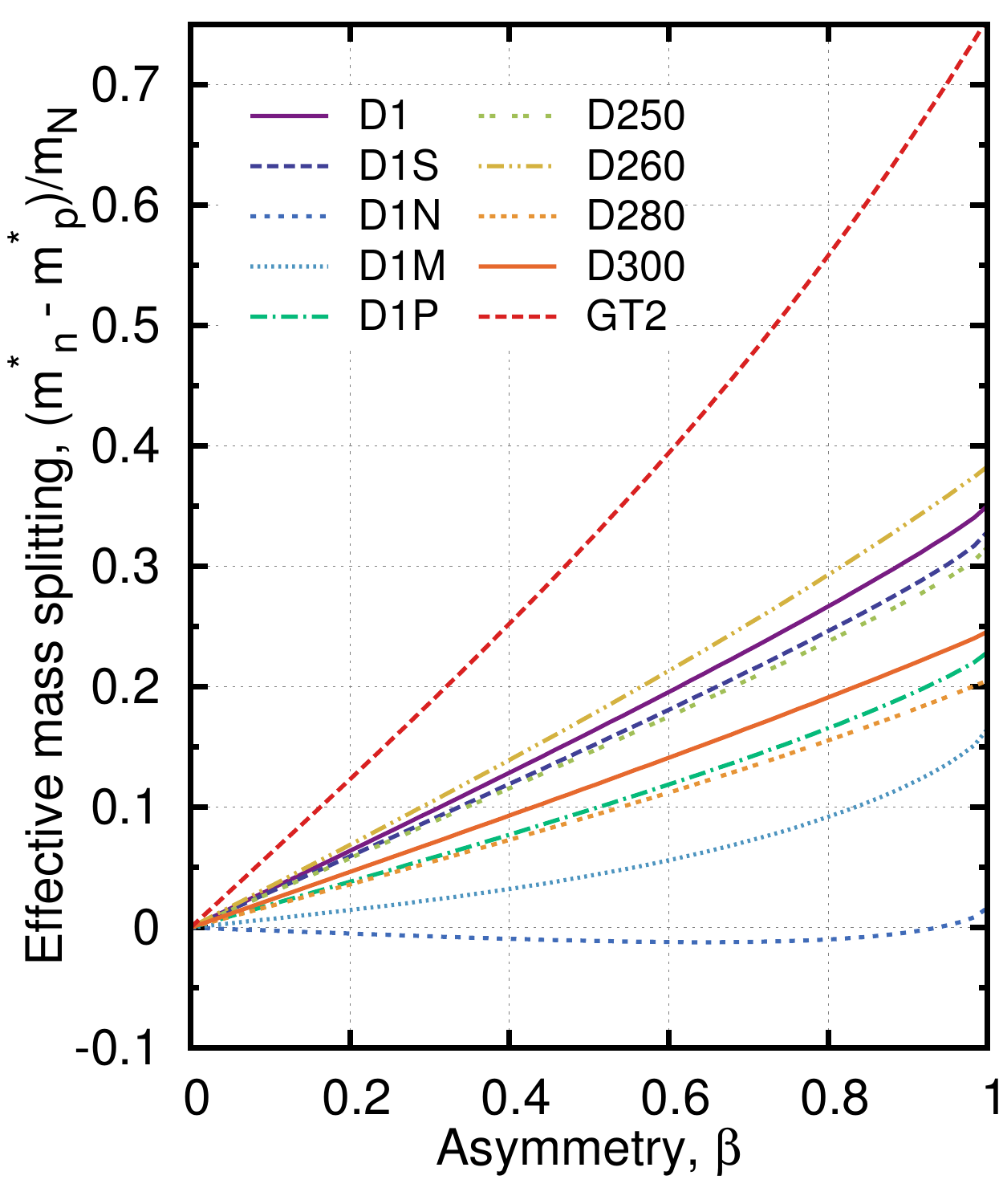}
\caption{(Color online) Isovector mass splitting at the Fermi surface, Eq.~(\ref{eq:isovector_mass}), as a function of isospin asymmetry at $\rho=0.16$ fm$^{-3}$. Most functionals show a linear dependence at small asymmetries.} 
\label{fig:isosplit_effmass}
\end{center}
\end{figure}

The effective mass is often characterised by its value at the Fermi surface, $m^*_\tau(k=k_F^\tau)$, rather than by its full momentum dependence. This allows for a simple characterisation of the variation of momentum dependence with asymmetry, at least close to the respective Fermi surfaces. Since $\mathsf{m}$ is symmetric in its two arguments, the isovector effective mass splitting is proportional to $B^i_{nn}$ only and to the difference of two symmetric $\mathsf{m}$  functions evaluated at the two Fermi surfaces:
\begin{align}
\label{eq:isovector_mass}
\frac{m_N}{m_n^*} - \frac{m_N}{m_p^*} =  
\sum_{i=1,2} B^i_{nn} & \left[ \mathsf{m} \left( \mu_i k_F^n, \mu_i k_F^n \right) \right.  \\
 &  \left. -\mathsf{m} \left( \mu_i k_F^p, \mu_i k_F^p \right) \right] \, .
\nonumber 
\end{align}
Because of its simplicity, this splitting can be analysed rather straightforwardly. For symmetric arguments, $\mathsf{m}(q_F,q_F)$ is a decreasing function of its argument up to $q_F \approx 2.38$. Consequently, as long as $ \mu_i k_F^{\tau} \lesssim 2.378$ and for $k_F^n > k_F^p$, the difference of $\mathsf{m}$ functions will be positive. In these conditions, the sign of the isovector splitting will be proportional to that of $B^i_{nn}$. 

These findings are confirmed by the results displayed in Figure~\ref{fig:isosplit_effmass}, where we show the isovector effective mass splitting as a function of isospin asymmetry at $\rho=0.16$ fm$^{-3}$. 10 functionals have positive splittings, with a neutron effective mass larger than a proton effective mass, $m^*_n>m^*_p$.  The exception is D1N, which has a very small but negative isovector splitting \cite{Chappert2007}. In contrast, since GT2 has a large and positive $B_{nn}^1$, its isovector mass splitting is almost twice as large as the results from other functionals. For the typical isospin asymmetry of heavy nuclei, $\beta = 0.2$, the isovector mass splitting predicted by most forces is of the order $\tfrac{m_n^*-m_p^*}{m_N} \approx 0.04-0.06$. If we identify this as the isovector $k-$mass splitting, this result is consistent with the findings of Ref.~\cite{Li2014}.

It is also interesting to note that most functionals show a linear dependence on $\beta$ for small asymmetries. This weak asymmetry dependence is entirely due to the shift in Fermi momenta within the $\mathsf{m}$ functions in Eq.~(\ref{eq:isovector_mass}). The slope of the curve could potentially be used to constrain $B^x_{nn}$, if accurate data for the isovector effective mass splitting was available. Efforts in this direction using optical model potentials are already underway \cite{Charity2014,*Charity2014a,Li2014}. A stronger dependence in asymmetry appears as the neutron matter limit is approached. At that point, the functional form of $\mathsf{m}$ becomes relevant and a non-linear shape is expected.

\section{Macroscopic properties}

\subsection{Isoscalar properties} 

In the previous section, we analysed the microscopic single-particle properties predicted by a variety of Gogny functionals. From now on, we concentrate on the results that these functionals predict for the bulk, thermodynamical properties. We start by analysing isoscalar and isovector properties at densities close to saturation. We present in Table~\ref{table:nm_properties} a series of properties of nuclear matter obtained by all the Gogny parametrizations of interest. We note that some of these properties, particularly in the isoscalar sector, are inputs in the fitting procedure. As such, they do not yield any new information. We provide the analytical expressions for these quantities in Appendix~\ref{app:macro}. We note that we have also employed numerical stencils to test our expressions. We crosschecked these data with the original publications as well as with other modern sources \cite{Margueron2001,Chappert2007,Than2009} and found a good general agreement.

\begin{table*}[t]
\begin{tabular*}{0.8\linewidth}{@{\extracolsep{\fill} } l | *{10}{c} }
\hline\noalign{\smallskip}
\hline\noalign{\smallskip}
Force & $\frac{m^*}{m}$ & $\rho_0$ & $e_0$ & $K_0$ &$Q_0$ & $S$ & $L$ & $K_\text{sym}$ & $Q_\text{sym}$ & $K_\tau$ \\
\hline
D1   & $   0.670$ & $   0.166$ & $  -16.30$ & $   229.4$ & $  -460.7$ & $   30.70$ & $   18.36$ & $  -274.6$ & $   616.7$ & $  -347.9$ \\
D1S  & $   0.697$ & $   0.163$ & $  -16.01$ & $   202.9$ & $  -515.8$ & $   31.13$ & $   22.43$ & $  -241.5$ & $   644.2$ & $  -319.1$ \\
D1N  & $   0.747$ & $   0.161$ & $  -15.96$ & $   225.6$ & $  -438.2$ & $   29.60$ & $   33.58$ & $  -168.5$ & $   440.1$ & $  -304.8$ \\
D1M  & $   0.746$ & $   0.165$ & $  -16.02$ & $   225.0$ & $  -459.0$ & $   28.55$ & $   24.83$ & $  -133.2$ & $   735.7$ & $  -231.6$ \\
D1P  & $   0.671$ & $   0.170$ & $  -15.25$ & $   254.1$ & $  -340.6$ & $   32.75$ & $   50.28$ & $  -159.3$ & $   408.3$ & $  -393.6$ \\
D1AS & $   0.670$ & $   0.166$ & $  -16.30$ & $   229.4$ & $  -460.7$ & $   31.30$ & $   66.55$ & $   -89.1$ & $   245.8$ & $  -354.7$ \\
D250 & $   0.702$ & $   0.158$ & $  -15.84$ & $   249.9$ & $  -362.0$ & $   31.57$ & $   24.82$ & $  -289.4$ & $   484.4$ & $  -402.4$ \\
D260 & $   0.615$ & $   0.160$ & $  -16.25$ & $   259.5$ & $  -373.1$ & $   30.11$ & $   17.57$ & $  -298.7$ & $   539.4$ & $  -378.9$ \\
D280 & $   0.575$ & $   0.153$ & $  -16.33$ & $   285.2$ & $  -298.5$ & $   33.14$ & $   46.53$ & $  -211.9$ & $   326.2$ & $  -442.4$ \\
D300 & $   0.681$ & $   0.156$ & $  -16.22$ & $   299.1$ & $  -237.5$ & $   31.22$ & $   25.84$ & $  -315.1$ & $   359.6$ & $  -449.6$ \\
GT2  & $   0.672$ & $   0.161$ & $  -16.02$ & $   228.1$ & $  -444.0$ & $   33.94$ & $    5.02$ & $  -445.9$ & $   741.2$ & $  -466.2$ 
\end{tabular*}
\caption{Isoscalar and isovector properties of nuclear matter as predicted by Gogny functional. All properties have units of MeV, except for the dimensionless effective mass, $\frac{m^*}{m}$ (column 2) and the saturation density $\rho_0$ in fm$^{-3}$ (column 3). \label{table:nm_properties}}
\end{table*}

The second column of Table~\ref{table:nm_properties} shows the symmetric matter effective masses obtained at the corresponding saturation densities (column 3). Note that the latter were obtained for each functional by using the same routines and underlying numerical constants, and hence the variability is entirely due to the Gogny functional parameters. The effective masses are computed at the Fermi surface, and hence they also correspond to the circles in the continuous lines of Fig.~\ref{fig:effm_tau}. In general, all effective masses fall within the range $m^* \approx ( 0.67\mbox{-}0.75)m$, which is slightly lower than modern Skyrme forces \cite{Chabanat1998} but in agreement with many-body estimates \cite{Jeukenne1976}. The exceptions are D260 ($m^*=0.615m$) and D280 ($m^*=0.575m$), which have lower effective masses. We note that these two functionals have $\alpha=\tfrac{1}{3}$, unlike D250 and D300 which were purposely designed with $\alpha=\tfrac{2}{3}$ \cite{Blaizot1995}. The range in compressibilities for a fixed $\alpha$ can only be achieved with substantial variations on the effective masses.

The equation of state (EoS) of symmetric nuclear matter is characterised by a few, relatively well-known parameters. The energy per particle presents a minimum around the saturation density, $\rho_0 \approx0.16$ fm$^{-3}$, at a value of $e_0 \approx -16$ MeV. The saturation densities and energies of the third and fourth columns of Table~\ref{table:nm_properties} mostly reflect the biases of the fitting input data, as reported in the original publications. In general, they are all close to the empirical values. We note, however, the relatively repulsive saturation energy and relatively large saturation density of D1P.\footnote{Our values for the effective mass, saturation density, energy, compressibility and symmetry energy for D1P differ slightly from the values in the original publication \cite{Farine1999}. This might be caused by a small difference in the value of $\alpha_1$. We use $\alpha_1=\tfrac{1}{3}$, but the results of the original publication are better reproduced with $\alpha_1 \approx 0.345$. These differences do not affect the results discussed in the following.} By construction, D1AS has the same isoscalar properties as D1 \cite{Ono2003}.

The curvature of the energy per particle around the saturation point can be described in terms of the incompressibility, $K_0$. This determines to a large extent the EoS of symmetric nuclear matter and is shown in the fifth column. An explicit expression for the incompressibility is given in Eq.~(\ref{eq:k0}). Most values are centered around the empirical estimate $K \approx 230$ MeV. The exceptions are, of course, the forces D250-D300, which were specifically obtained to match a given compressibility. We note, in particular, that D280 is slightly off its expected value. A further characterisation of the saturation point is obtained from the skewness, $Q_0$ (see Eq.~(\ref{eq:q0}) for an explicit expression). In contrast to $\rho_0$, $e_0$ and $K_0$, the skewness is not generally included in functional fits. Consequently, its value is not as constrained. We find  that Gogny functionals predict a negative skewness, in the range between $Q_0 \approx -237$ and $Q_0 \approx -516$ MeV. This agrees with a recent study based on a combined analysis of flow data and neutron star masses in the relativistic mean-field picture, which suggests $-494 < Q_0 < -10$ MeV \cite{Cai2014}. Note that neither the compressibility nor the skewness receive a contribution from $A_0^i$. $K_0$ and $Q_0$ are therefore entirely determined by a competition between the zero-range and the finite-range exchange terms (in addition to the sizeable kinetic terms, of course).

\begin{figure}[t!]
\begin{center}
\includegraphics[width=0.7\linewidth]{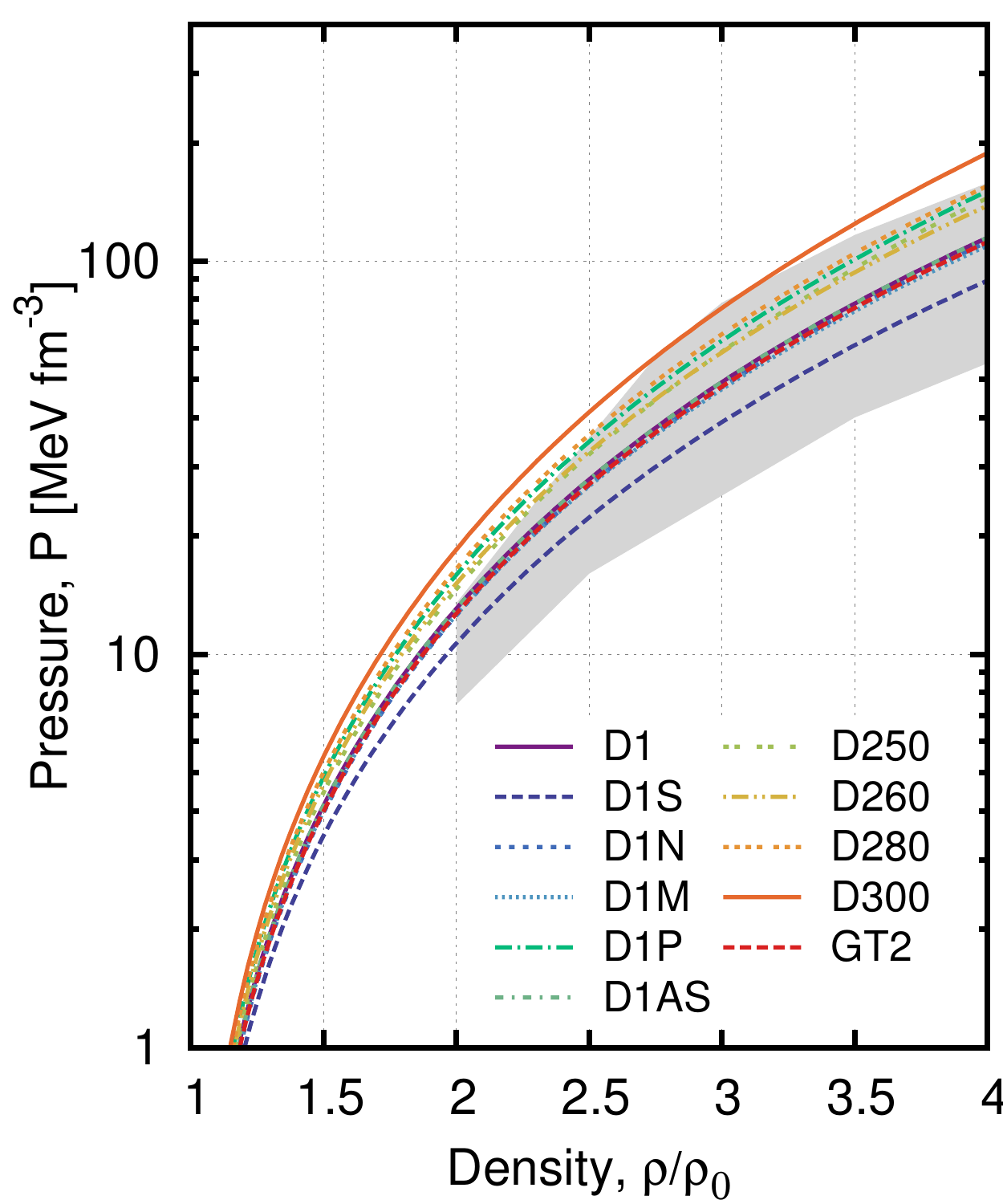}
\caption{(Color online) EoS (pressure as a function of density) for symmetric nuclear matter for all Gogny functionals. The density is divided by the respective saturation density of each functional. The shaded region corresponds to the EoS consistent with the experimental flow data of Ref.~\cite{Danielewicz2002}.}
\label{fig:EoS_danielewicz}  
\end{center}
\end{figure}

The saturation energy minimum is reflected in a zero for the pressure of symmetric matter, $P_\text{SNM}$. We provide an explicit expression for this function in Eq.~(\ref{eq:psnm}). Because of the positive compressibility, one expects a positive and increasing pressure as a function of density. We confirm that this is indeed the case in Fig.~\ref{fig:EoS_danielewicz}, which shows the symmetric matter EoS for the 11 Gogny functionals. We have concentrated on densities well above saturation, where the constraints coming from experimental flow data in intermediate energy heavy-ion collisions are relevant \cite{Danielewicz2002}. Some of the Gogny predictions had already been compared to flow results in Ref.~\cite{Than2009}, but ours is a  more comprehensive analysis. 

First, we stress the fact that all Gogny parametrizations produce relatively similar results for the pressure at high densities. This is in stark contrast to the isovector properties presented in the following subsection. Together with the results of Table~\ref{table:nm_properties}, this indicates that systematic uncertainties, \emph{i.e.} those due to differences in the functionals, are rather small in the isoscalar sector. In other words, the EoS of symmetric matter is under good control, as one might expect from the fitting procedure of the functionals. Note that the combinations of isoscalar $A_0$ and $C_0$ matrix elements are relatively well constrained (see Fig.~\ref{fig:matrix_elements}) compared to the finite-range exchange elements $B_0$. All these elements enter the expressions for the energy, Eq.~(\ref{eq:esnm}), pressure, Eq.~(\ref{eq:psnm}), and compressibility, Eq.~(\ref{eq:k0}), and hence can't be disentangled easily in a fitting procedure that includes these as input.

Second, most functionals fall within the experimental constraints obtained from flow data \cite{Danielewicz2002}. The differences between equations of state can be succinctly explained by the compressibility of each functional. If we order the functionals in terms of the value of $K_0$, the functional with a largest compressibility, D300, produces the largest pressure at high densities. Consequently, the EoS falls above the empirical constraints at all densities. The following subset, with $K_0 \approx 250-280$ MeV, is formed of D280, D260, D1P and D250. Their pressures are only consistent with the flow data above $\rho \gtrsim 2.5 \rho_0$. Most other functionals were fitted with $K_0 \approx 230$ MeV and hence their equations of state are very close to each other. D1S, which has the lowest compressibility, yields the lowest pressures, but is still well within the flow data. We note that, as long as $K_0 < 240$ MeV, the EoS is compatible with the experimental flow data. 
We note, however, that the flow data of Ref.~\cite{Danielewicz2002} was obtained with a mean-field with a momentum dependence characterised by $m^* \approx 0.7 m$ and a medium-modified nucleon-nucleon cross section. The full momentum dependence of Gogny functionals might in principle affect these results. A further characterisation of the pressure in terms of skewness lies beyond the scope of this work.

\subsection{Isovector properties at saturation} 

The characterisation of isovector bulk properties for Gogny functionals is one of the main aims of this work. We provide the values for some macroscopic isovector properties, computed at saturation, in columns 7-11 of Table~\ref{table:nm_properties}. The symmetry energy, $S$, has been fitted in some of the functionals (D1 \cite{Decharge1980}, D1AS \cite{Ono2003}, D1M \cite{Goriely2009}). In others, it is indirectly constrained by fits to the neutron matter equation of state (D1P \cite{Farine1999}, D1N \cite{Chappert2008}). In contrast, fitting procedures for functionals like the D250-D300 family  \cite{Blaizot1995} have not explicitly considered isovector bulk properties. In spite of these differences, we find that the values of the saturation symmetry energy are generally within the range $S\approx 30-34$ MeV. We note that D1M has a particularly low value for the symmetry energy, imposed during the fitting procedure \cite{Goriely2009}. 

The slope parameter of the symmetry energy, 
\begin{align}
\label{eq:lief}
L(\rho) = 3 \rho \frac{ \partial S(\rho)}{\partial \rho} \, ,
\end{align}
quantifies the density dependence of the symmetry energy. It is particularly relevant for neutron star physics, since it determines the pressure of neutron matter close to saturation \cite{Piekarewicz2009}. We provide the values of $L$ at saturation in column 8 of Table~\ref{table:nm_properties}. The variation in values of $L$ is extremely wide, and we find that most Gogny functionals are outside the empirically constrained ranges. More details are provided below, but let us highlight the extremely low values of $L\approx5$ MeV for GT2 and $L \approx18$ MeV for both D1 and D260. Inevitably, this will have a negative impact on the neutron star properties associated with these functionals. 

\begin{figure}[t!]
\begin{center}
\includegraphics[width=\linewidth]{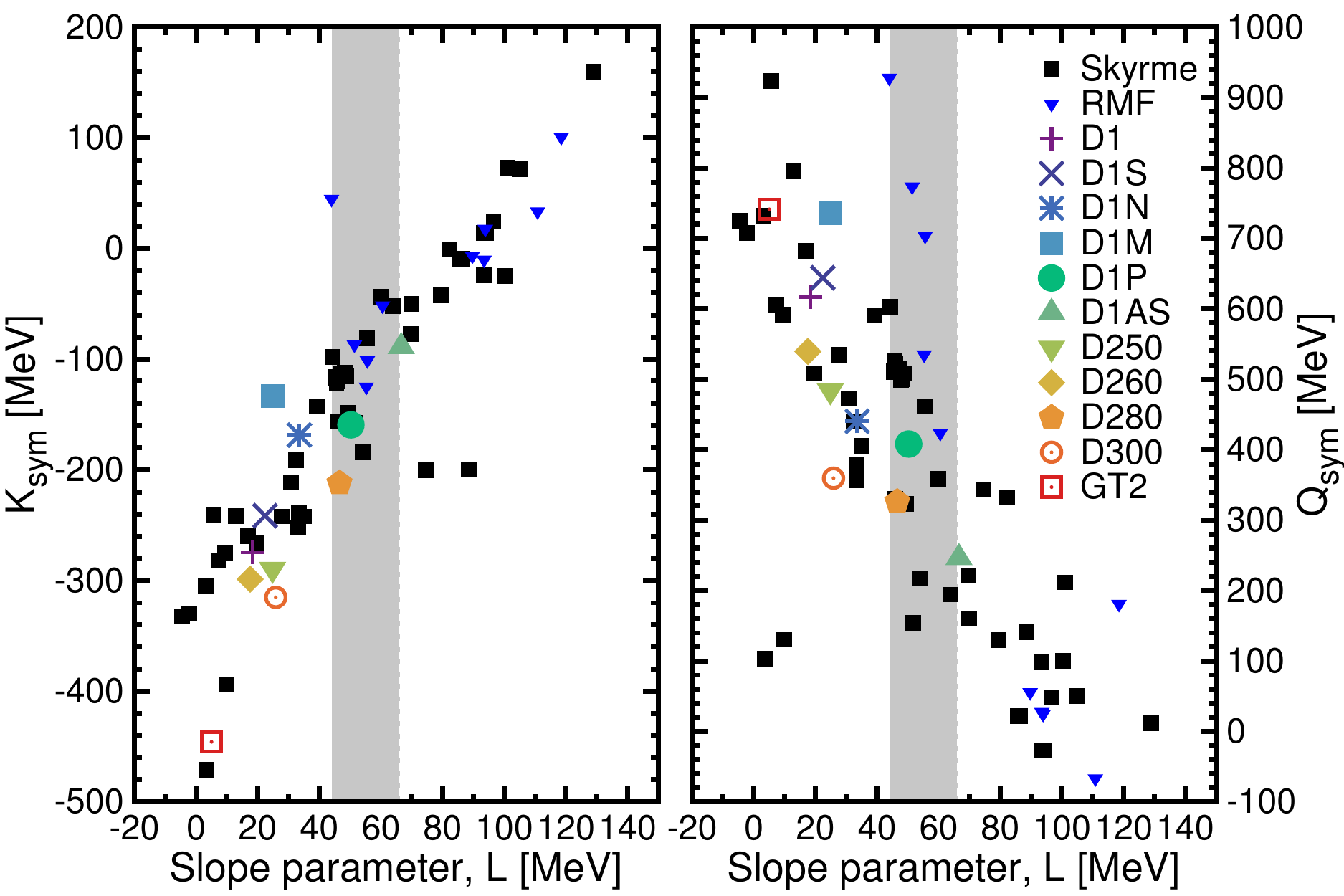}
\caption{(Color online) $K_\text{sym}$ (left panel) and $Q_\text{sym}$ (right panel) as a function of the slope parameter $L$ for all Gogny functionals. Small black squares (small triangles) are representative of Skyrme (relativistic mean field) functionals. The shaded region corresponds to the constraints obtained by Lattimer and Steiner in Ref.~\cite{Lattimer2014}.}
\label{fig:ksym_qsym}
\end{center}	 
\end{figure}

Columns 9 and 10 of Table~\ref{table:nm_properties} show the values of $K_\text{sym}$ and $Q_\text{sym}$ at saturation for all the functionals. These properties are less constrained by empirical data. In general, we find negative large values of $K_\text{sym}$, in the range $-90$ to $-446$ MeV. These are far more negative than the values predicted by microscopic calculations \cite{Vidana2009}. Similarly, Gogny functionals predict positive and relatively large values of $Q_\text{sym}$, in contrast to the negative and small values obtained by Brueckner--Hartree--Fock calculations \cite{Vidana2009}. 
We find the expected general trends obtained in other mean-field approaches \cite{Chen2009,Centelles2009}. We show in Fig.~\ref{fig:ksym_qsym} the values of $K_\text{sym}$ (left panel) and $Q_\text{sym}$ (right panel) as a function of the slope parameter. Gogny functionals with lower values of $L$, like GT2, D300, D260 or D1, tend to have more negative values of $K_\text{sym}$ and more positive values of $Q_\text{sym}$. These follow the generic correlations of Skyrme and relativistic mean-field calculations, which we show in small squares and triangles in the Figure \cite{Vidana2009}. All in all, this suggests that the correlations arise from basic underlying isovector physics. For the Gogny functionals, the correlations must arise from a few of the underlying parameters, as shown in the expressions of Eqs.~(\ref{eq:lrho})-(\ref{eq:qsymrho}). Note, in particular, that both $K_\text{sym}$ and $Q_\text{sym}$ are independent of the isovector direct finite-range matrix elements, $A_1^i$.

The asymmetry dependence of the compressibility is determined by a parameter $K_\tau$ which, to lowest order, is given by the combination $K_\tau \equiv K_\textrm{sym} - 6 L - (Q_0/K_0) L$ \cite{Piekarewicz2009,Chen2009}. This asymmetry dependence can potentially be extracted from Giant Monopole Resonance experiments \cite{Patel2012}. The value $K_\tau = -550 \pm 100$ MeV is nowadays generally accepted \cite{Li2007}. Most Gogny forces, as found in column 11 of Table~\ref{table:nm_properties}, sit somewhat above the less negative end, with values between $K_\tau \approx -450$ and $-300$ MeV. As a matter of fact, taking this experimental value seriously, only GT2 and D300 would be valid, in spite of their very low slope parameters. We note that Gogny functionals indicate a preference for less negative values of $K_\tau$, in agreement with the theoretical analysis presented in Ref.~\cite{Centelles2009} as well as with the correlation analysis of Ref.~\cite{Chen2009}. We also indicate that $K_\tau$ is only the lowest order term in the asymmetry dependence of the compressibility, and higher-order contributions can be relevant \cite{Chen2009}. Further research on finite nuclei properties, particularly resonances, with Gogny interactions will provide a further consistency check \cite{Peru2014}. 

\begin{figure}
\begin{center}
\includegraphics[width=0.7\linewidth]{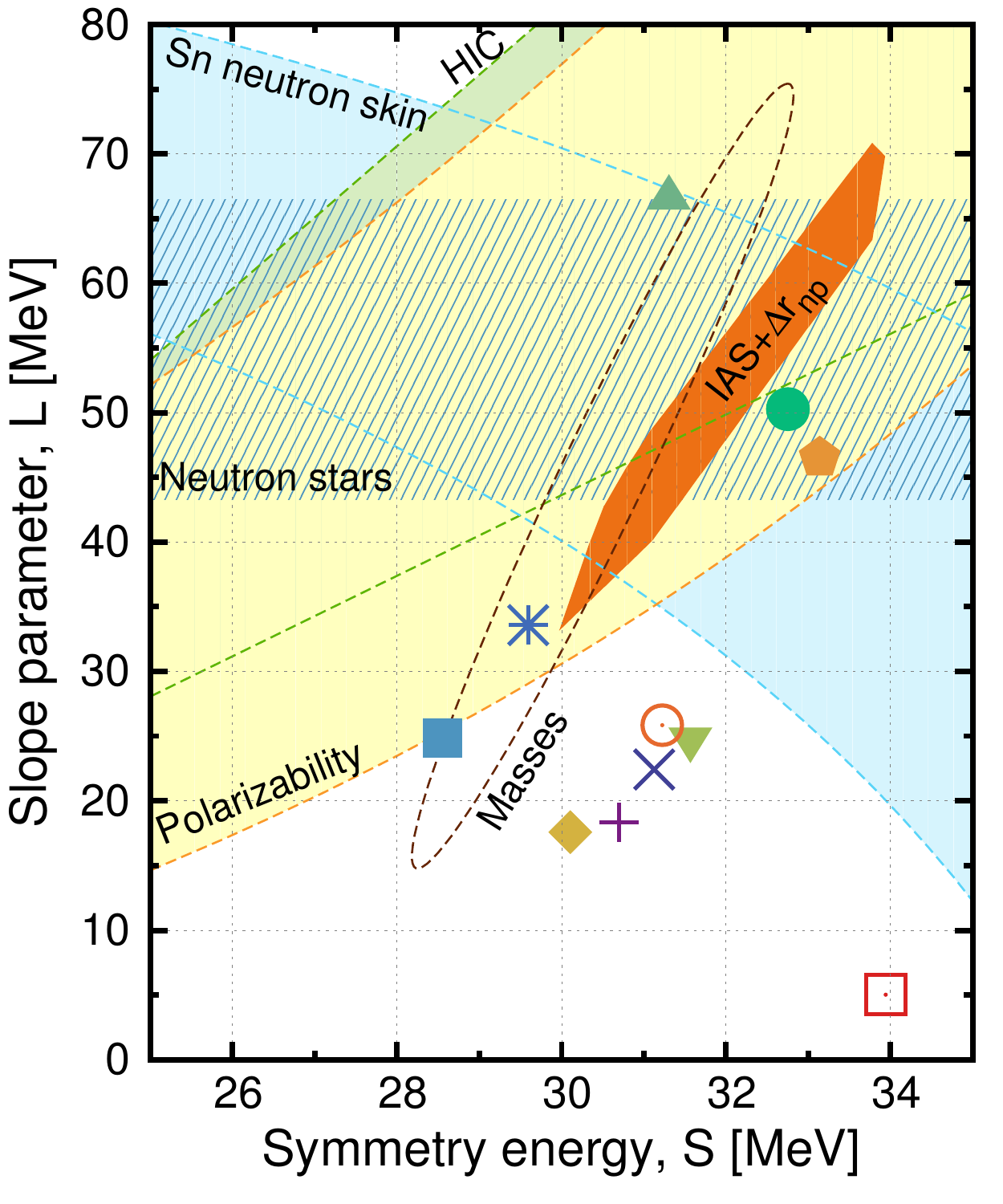}
\caption{(Color online) Slope parameter as a function of symmetry energy  for all Gogny functionals. The symbols show the values at saturation, whereas lines correspond to the evolution with density of both parameters (see Fig.~\ref{fig:esym}) for a legend. The regions corresponds to empirical constraints, following Refs.~\cite{Lattimer2013,Lattimer2014}}
\label{fig:esym_lpar}
\end{center}
\end{figure}

In contrast to the relatively poorly constrained $K_\tau$, there is an abundance of empirical evidence that helps restrict the values of the saturation symmetry energy and its slope. These two properties are generally tightly correlated, so their independent determination is difficult \cite{Dobaczewski2014,Goriely2014}. We show in Figure~\ref{fig:esym_lpar}  the values of $S$ and $L$ for all Gogny functionals under consideration. In addition, following Refs.~\cite{Lattimer2013,Lattimer2014}, we show some of the empirical constraints obtained from a variety of methods. The $68 \, \%$ confidence ellipse labeled ``Masses" is obtained by propagating the fit errors in a density functional calculation using the UNEDF0 functional \cite{Kortelainen2010}, with the choice $\sigma=1$ MeV \cite{Lattimer2014}. The Sn neutron skin thickness results come from the analysis of Chen and co-authors \cite{Chen2010}. 
From this analysis, one can also fit a relation between the skin thickness in lead and the corresponding symmetry energy and slope parameters of a variety of functionals. In addition, there is a tight linear correlation between the dipole polarizability and the skin thickness \cite{Roca-Maza2013}. If we use the fit parameters of Ref.~\cite{Lattimer2013} for the former and of Ref.~\cite{Roca-Maza2013} for the latter, we find the ``polarizability" band in the figure. Note that the position and the width of the band can change if different fits are used for either correlation. 

Isospin diffusion studies in heavy-ion collisions, labelled HIC, provide additional constraints in the region of small symmetry energies \cite{Tsang2009}. The band labelled ``neutron stars" is obtained by fconsidering the $68 \, \%$ confidence values for $L$ obtained from a Bayesian analysis of simultaneous mass and radius measurements of neutron stars \cite{Steiner2013}. Finally, a narrow and small diagonal region above $S>30$ MeV is obtained from simultaneous constraints of Skyrme--Hartree--Fock calculations of isobaric analog states (IAS) and the $^{208}$Pb neutron skin thickness \cite{Danielewicz2014}. 

Strikingly, 6 of the 11 Gogny parametrizations fall outside of all empirical determinations of $S$ and $L$. In all cases, the value of $L$ is below the expected results. D1N and D1M sit within the polarizability and the masses constraints, but their slopes are still small compared with the neutron stars constraints.  Only D1P, D1AS and D280 have large enough slopes to fit within the polarizability, neutron skin and astrophysical constraints. As for $S$, these functionals predict values within $1$ MeV of the average value of all observations, $S \approx 31$ MeV. We note that no parametrization falls within the IAS+$\Delta r_{np}$ region or within the joint constraint region. 

Fig.~\ref{fig:esym_lpar} demonstrates graphically one of the main conclusions of this work. Few, if any, Gogny functionals show a good simultaneous reproduction of the symmetry energy, $S$, and its slope, $L$, at saturation as of the present empirical constraints. An analysis of the density dependence of these quantities in the following subsection  will demonstrate that this is largely a consequence of the poor constraints on the isovector finite-range part of the functional. Future fits of Gogny functionals risk lacking quality in the isovector sector unless these properties are fit consistently. We point out, however, that the comparison with empirical constraints should be performed with some caution. Most of these constraints were obtained by using zero-range functionals of the Skyrme type \cite{Kortelainen2010,Chen2010,Roca-Maza2013,Danielewicz2014}. A consistent determination of the correlations depicted in the Figure using finite-range functionals is still missing. The last section of this paper, where neutron stars are analysed, is an effort in this direction. 

\subsection{Density dependence of isovector properties} 

\begin{figure}[t!]
\begin{center}
\includegraphics[width=0.7\linewidth]{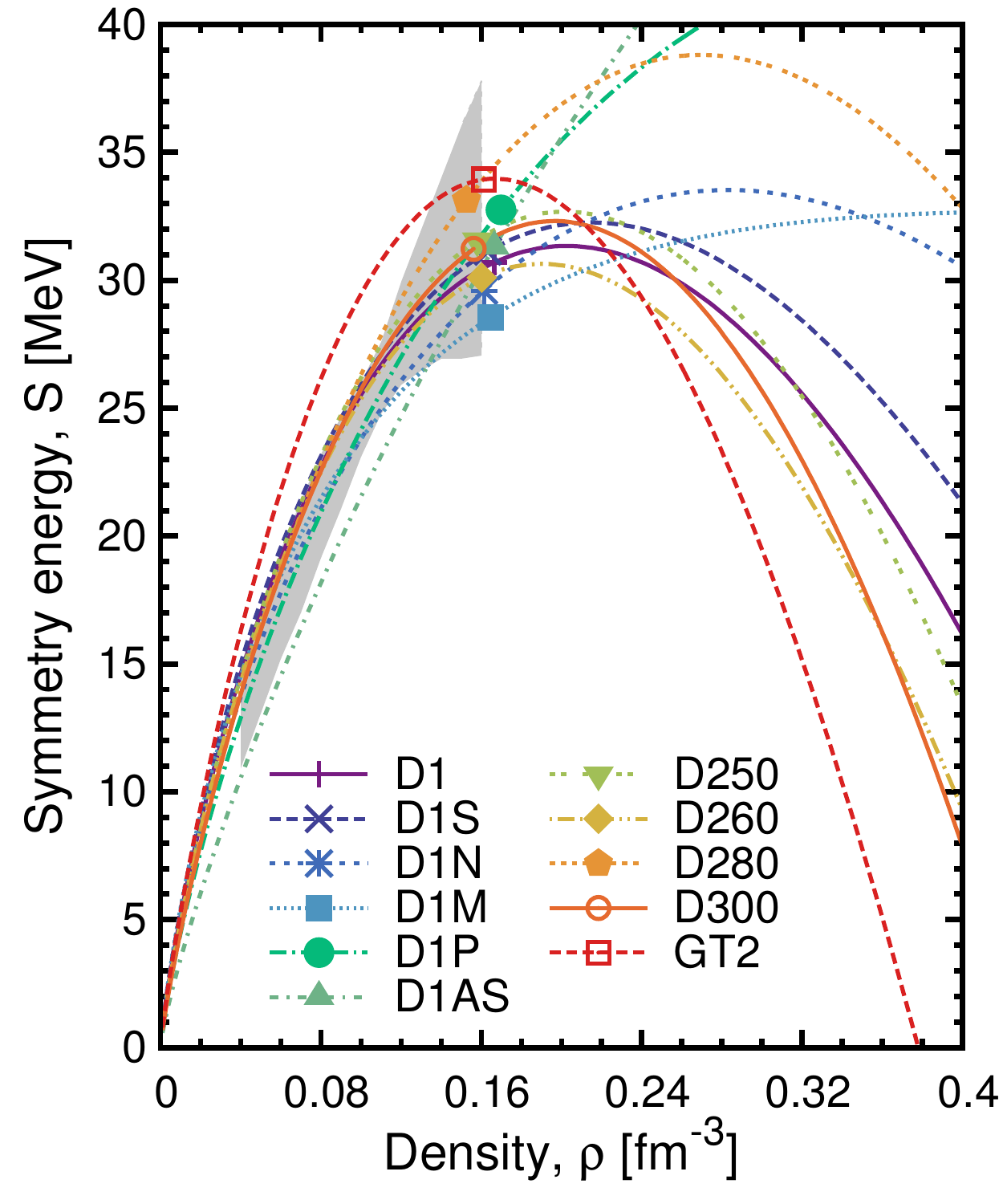}
\caption{(Color online) Symmetry energy as a function of density for all Gogny functionals. The shaded region corresponds to the constraints arising from IAS of Ref.~\cite{Danielewicz2014}.}
\label{fig:esym}
\end{center}	 
\end{figure}

In addition to the values at saturation, there have been a wide range of attempts to determine the density dependence of the symmetry energy and, generally speaking, of all isovector properties. With Gogny functionals, an analytic expression of the symmetry energy, $S(\rho)$, is provided in Eq.~(\ref{eq:srho}). Note that, in addition to the standard isovector zero- and direct finite-range terms, the exchange finite range contribution provides a non-trivial density dependence. The density dependence of the symmetry energy is explored in Fig.~\ref{fig:esym}. We show results for all Gogny functionals and highlight the value of $S$ at saturation. These results are compared to the constraints from the IAS analysis with Skyrme functionals \cite{Danielewicz2014}.

The sub-saturation region has a relatively small systematic error. Most functionals fall within or very close to the empirically constrained region. GT2, however, has a relatively large symmetry energy below $\rho_0$ and overshoots the constraints in this region. In contrast, D1AS predicts relatively low symmetry energies for $\rho<0.14$ fm$^{-3}$, but it has a much stiffer density dependence above $\rho_0$. The fact that most Gogny functionals reproduce the low-density symmetry energy indicates that it is well-constrained by finite nuclei data. 

The largest discrepancies between functionals are observed above saturation density, as expected. We stress the fact that  most functionals show either a maximum or a plateau in the symmetry energy as a function of density and that this occurs, in most cases, right above saturation. As a consequence, the values of the slope $L$ are relatively small for most functionals. The exceptions are D1AS and D1P, which have a rather stiff character. Above saturation, several functionals display a sharp decrease in density, which will eventually lead to negative values of $S$ at densities beyond the range displayed in Fig.~\ref{fig:esym}. For GT2, the change in sign of the symmetry energy happens already around $\rho \approx 0.38$ fm$^{-3}$. The presence of this isospin instability would have consequences in both neutron stars \cite{Margueron2001,Margueron2002,Goriely2010} and heavy ion collisions \cite{Li2002}. Note, however, that microscopic calculations do not predict any signs of such a transition \cite{Vidana2009}.

\begin{figure}[t!]
\begin{center}
\includegraphics[width=0.7\linewidth]{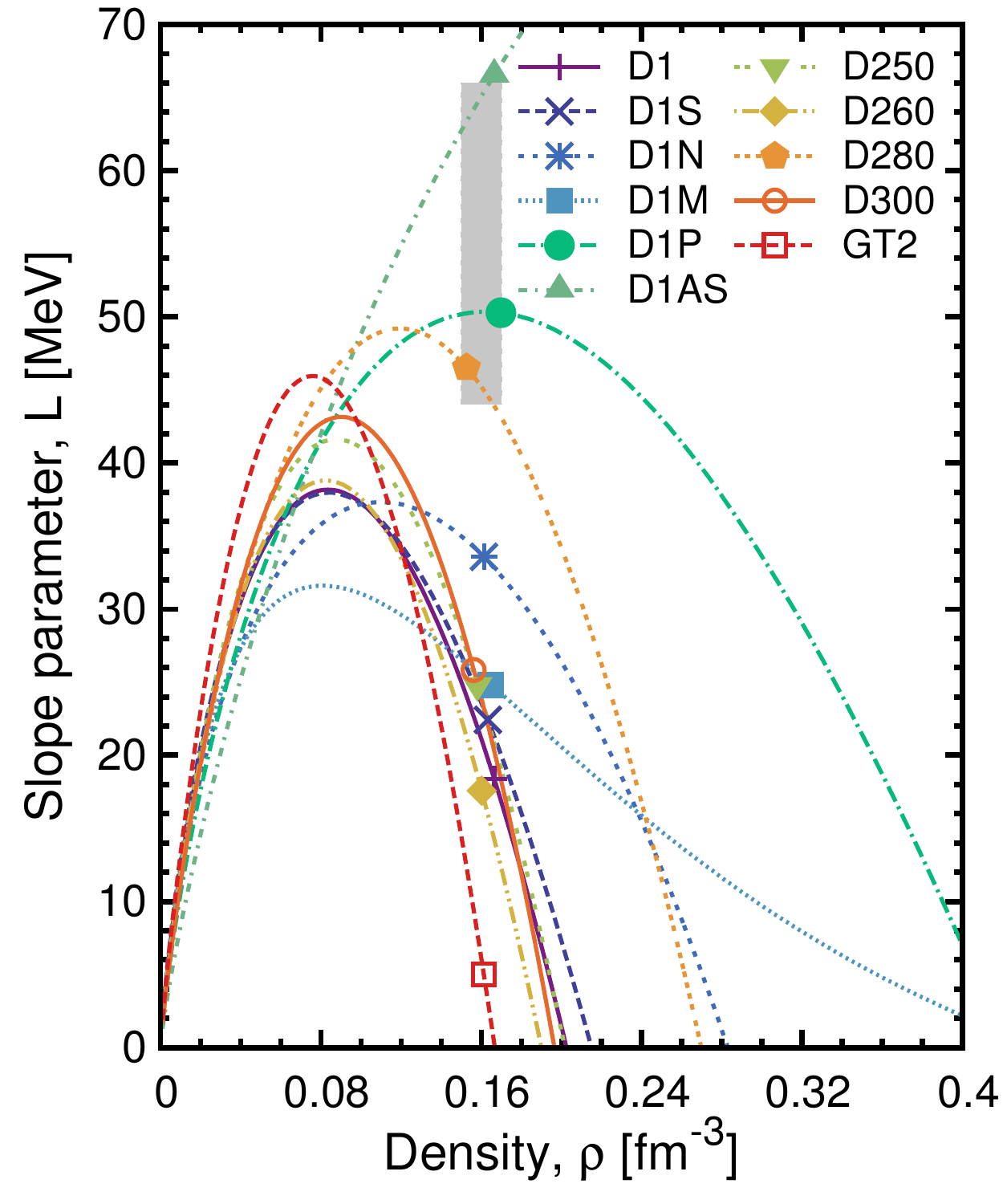}
\caption{(Color online) Slope parameter as a function of density for all Gogny functionals. The shaded region corresponds to the combined constraints obtained by Lattimer and Steiner in Ref.~\cite{Lattimer2014}.}
\label{fig:lpar}
\end{center}
\end{figure}

A complementary perspective of these results is obtained by analysing the density dependence of the slope parameter, $L$. We provide an expression for this quantity in Eq.~(\ref{eq:lrho}). Fig.~\ref{fig:lpar} shows the density dependence of the slope parameter for all Gogny functionals. We also display the ``neutron star" band, $L \approx 44-66$ MeV, discussed already in the context of Fig.~\ref{fig:esym_lpar}. Most Gogny functionals show a peak and a subsequent decrease in $L$ around half saturation density. Consequently, the value of $L$ at saturation becomes relatively small, falling below the empirical estimates. One therefore expects Gogny forces to predict rather small neutron skin thicknesses in $^{208}$Pb. This is the case for both D1S and D1N, as observed in Ref.~\cite{Roca-Maza2011}. More importantly, the pressures for isospin asymmetric matter as predicted by these functionals will be smaller than empirical estimates suggest. The structure of neutron-rich isotopes will be affected by the unrealistic prediction of slopes. Note also that about half the functionals predict negative slopes within $0.05$ fm$^{-3}$ of around saturation, in accordance with the very soft symmetry energies observed around saturation in Fig.~\ref{fig:esym}. It is worth mentioning that D1M shows a rather unique density dependence for the slope, with a nuanced decrease in density above saturation. For this functional, $L$ does not become negative below $\rho > 0.42$ fm$^{-3}$. 

Three functionals fall within the empirical range predicted by neutron star physics. For D280 and D1P, which have large enough values of $L$ at saturation, the slope is close to a maximum at saturation. The decrease of $L$ with density eventually leads to decreasing symmetry energies at relatively large densities ($0.26$ fm$^{-3}$ for D280 and $0.42$  fm$^{-3}$  for D1S). In contrast, D1AS is the only functional predicting a monotonically increasing slope parameter - and a consequently stiff symmetry energy. We note that this was achieved by construction, adding a density-dependent zero-range term to the functional \cite{Ono2003}. Because of their large pressures for neutron-rich matter around saturation, one would expect relatively large neutron stars associated with these functionals. We will confirm this tendency in the following section.  

\begin{figure*}[t!]
\begin{center}
\includegraphics[width=0.7\linewidth]{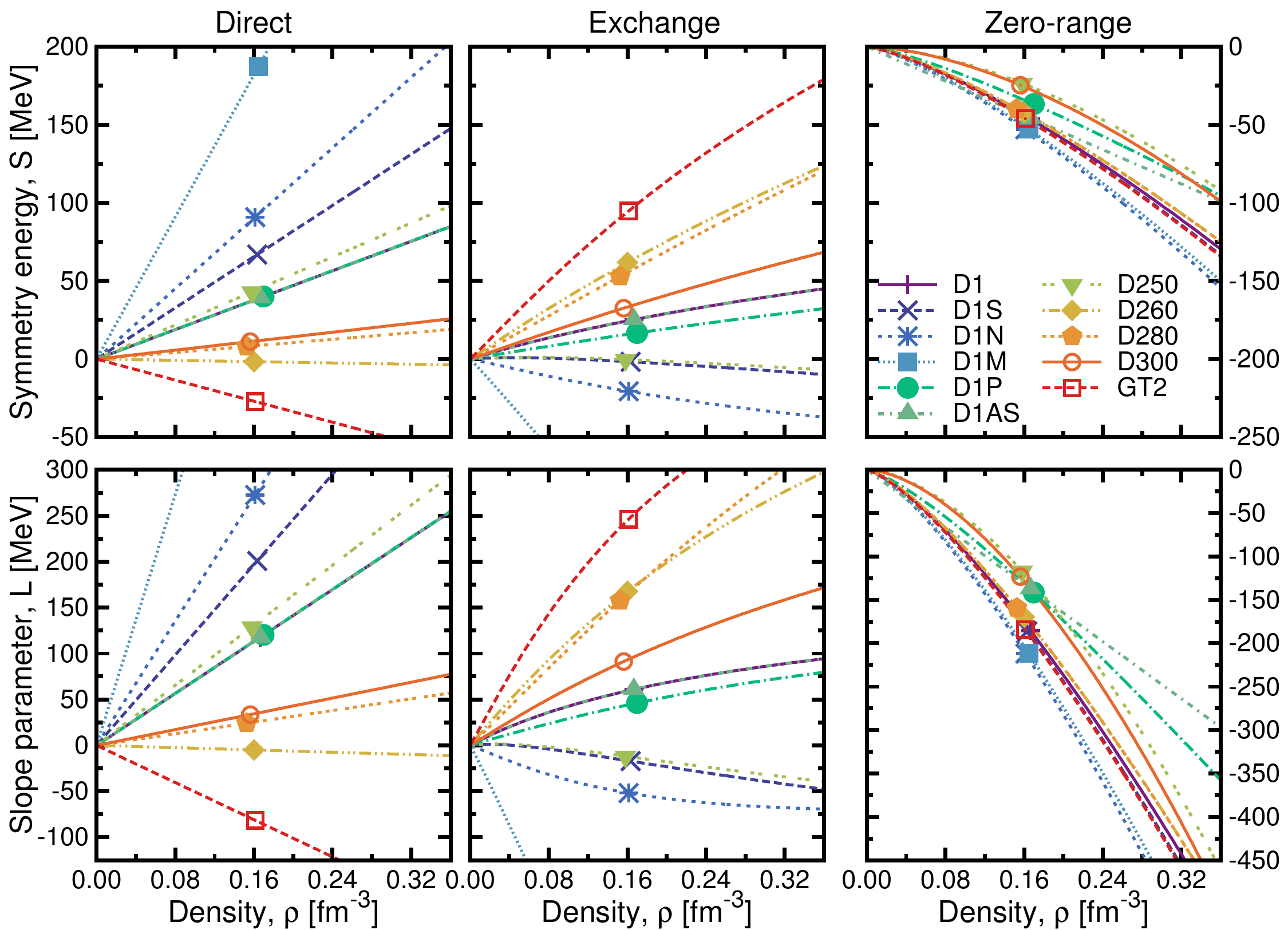}
\caption{(Color online) Direct (left panels), exchange (central panels) and zero-range contributions (right panels) of the symmetry energy (top panels) and the slope parameter (bottom panels). Results for 11 functionals are shown. The symbols correspond to the respective saturation points. Note that the left and central panel share the same $y$ axis.}
\label{fig:esym_lpar_components}
\end{center}
\end{figure*}

What is the underlying cause for the rapid decrease of the symmetry energy with density in most Gogny functionals? One can attempt to answer this question by separating different contributions to both the symmetry energy and the slope parameter. In Fig.~\ref{fig:esym_lpar_components}, we provide the contributions to both quantities arising from the direct finite-range term, the exchange finite-range term and the zero-range term (which includes both direct and exchange at once). The direct contribution to both quantities is particularly simple, as seen from Eqs.~(\ref{eq:srho}) and (\ref{eq:lrho}):
\begin{align*}
	S_\text{direct}(\rho) = \frac{1}{2} \sum_{i=1,2} A_1^i \rho \, .
\end{align*}
Note, in particular, that the term is linear in density and that only the sum of all isovector matrix elements plays a role. Because the values of these matrix elements are relatively widely spread (see Fig.~\ref{fig:matrix_elements}), we expect a relatively large systematic uncertainty. This is precisely what we find in the left panels of Fig.~\ref{fig:esym_lpar_components}. D1M has extremely large and positive contributions. In contrast, GT2 provides a negative contribution as density increases. Because the combination $\sum_i A_1^i$ is very similar for some functionals, we find that D1P (and D250 to a lesser extent) and D1 have very similar values of $S_\text{direct}$ and $L_\text{direct}$. The direct contribution to the slope parameter (bottom left panel) is  $3$ times that of the symmetry energy (top left panel), $L_\text{direct}(\rho) = 3 \times S_\text{direct}(\rho)$. Note also that, as expected, D1 and D1AS have the same direct finite-range contributions. 

The exchange contribution to the isovector properties (central panels of Fig.~\ref{fig:esym_lpar_components}) shows a density dependence that is, to a certain extent, the opposite of the direct term. GT2, for instance, is now large and positive, whereas D1M is large and negative. In general, these terms increase substantially with density, even though the dependence is not linear anymore. In fact, the density dependence is dictated by non-trivial functions of the Fermi momenta, see Eqs.~(\ref{eq:s12}) and (\ref{eq:l12}). With these equations, it is easy to show that in the low density limit the exchange contributions to $S$ and $L$ are (a)  linear in  density, (b) proportional to $B_i^1$ only and (c) related by $L_\text{exc}(\rho) = 3 \times S_\text{exc}(\rho)$. The exchange contributions are of the same order as the direct term, which indicates that large cancellations are necessary to get isovector properties of a natural size.

The right panels of Fig.~\ref{fig:esym_lpar_components} show the density dependence of the zero-range contributions to the isovector properties. First, we note that the dependence on the functional is largely reduced. The fitting procedure of Gogny functionals is such that part of the zero-range, density-dependent term is often fixed from the start.\footnote{For all functionals, $\alpha_1$ is chosen to be $\tfrac{1}{3}$ or $\tfrac{2}{3}$. $x_0$ is also often fixed, which leaves very little room for $t_0^1$ to change substantially. } Consequently, the parameter space for this term is substantially reduced and $S_\text{ZR}(\rho)$,
\begin{align}
	S_\text{ZR}(\rho) = \frac{1}{2} \sum_{i=1,2} C_1^i \rho^{\alpha_i+1} \, ,
\end{align}
 is very similar for all functionals. Also, because in most parametrizations $C_1^2=0$, the relation, 
\begin{align}
	L_\text{ZR}(\rho) = 3 \left( \alpha_i +1 \right) S_\text{ZR} (\rho) \, ,
\end{align}
holds. 

Second, and most important, the zero-range density-dependent term is negative at all densities. In the low density limit, the linear density dependence of both the direct and the exchange dominates (although absolute values are determined by large cancellations between terms). In the limit of large densities, in contrast, Eqs.~(\ref{eq:s12}) and (\ref{eq:l12}) suggest that the exchange term is proportional to $k_F \approx \rho^{1/3}$. In addition, for most functionals the zero-range term has a power $\alpha_1=\tfrac{1}{3}$ (and $\alpha_2=0$) which therefore involves $S_\text{ZR} \to \sum_i C_1^i \rho^{4/3}$. Together with the linear dependence on the direct term, it appears that the zero-range term inevitably dominates the density dependence of isovector properties at high densities. As a consequence, the symmetry energy of most parametrizations becomes soft at relatively low densities. We note that D1P is the only functional with a second zero-range term, that is, $t^2_0 \neq 0$. It performs well in the isovector sector.
Future parametrizations could use this freedom to improve the density dependence of isovector properties. Specifically, the fact that $S_\text{exc}$ is not proportional to $L_\text{exc}$ for $t_0^2 \neq 0$ can help break the tension between these two isovector parameters. 

The decomposition in terms of direct, zero-range, and exchange terms of the isovector properties is arbitrary to a large extent. Other decompositions can be used to pinpoint similar issues. In Ref.~\cite{Chen2012}, Chen and coauthors propose a split of the symmetry energy and the slope in terms of contributions arising from the single-particle potential at the Fermi surface, based on the Hugenholtz-van Hove theorem. We  performed an extensive study of this decomposition, but we present here only a summarised version. Within this approach, the symmetry energy can be decomposed into two terms:
\begin{align}
	S_1(\rho) &= \frac{1}{3} \frac{ \hbar^2 k_F^2}{ 2 m_0^*(\rho)}\, , \\
	S_2(\rho) &= \frac{1}{2} U_1^\text{sym}(k_F) \, .
\end{align}
The first term is a non-locality corrected kinetic contribution. The non-local correction, evidenced by the effective mass, is of the order of $\approx 30 \, \%$ at saturation and usually increases with density. The competition between the Fermi momentum and the effective mass can give rise to $S_1$ terms that decrease with density. The second term, $S_2$, is proportional to the value of the isovector single-particle potential at the Fermi surface. As observed in Fig.~\ref{fig:u_iso_sc_vec}, the value of $U_1$ at the Fermi surface can vary substantially. For quite a few functionals, this has a tendency to become negative slightly above saturation and drives the decrease of $S$ with density \cite{Chen2012}. A similar decomposition exists for the slope parameter, but it involves 5 different terms. There is in general a substantial Gogny functional dependence of these terms. This is particularly important for $L_4$ and $L_5$, which are associated with higher orders in the single-particle potential expansion \cite{Chen2012}. $L_4$ and $L_5$ are entirely due to the finite-range exchange part of the functional.\footnote{$L_5$ also receives a contribution from the rearrangement potential, associated with the density-dependent zero-range term.}

\section{Neutron matter and neutron stars}

Neutron matter provides a particularly sensitive test to the isovector behaviour  of nuclear energy density functionals \cite{Lattimer2012}, which is interesting because it can be straightforwardly connected to neutron stars \cite{Erler2013}. Present and future generation observations will eventually constrain the mass-radius relation for astrophysical compact objects, and this information can be fed back into the neutron matter EoS \cite{Steiner2010}. Even before that happens, essential tests regarding the consistency and isospin dependence of energy density functionals should be performed. We note that this task has been already carried out in extensive studies of Skyrme functionals \cite{Stone2003,Dutra2012}. 

\begin{figure}
\begin{center}
\includegraphics[width=\linewidth]{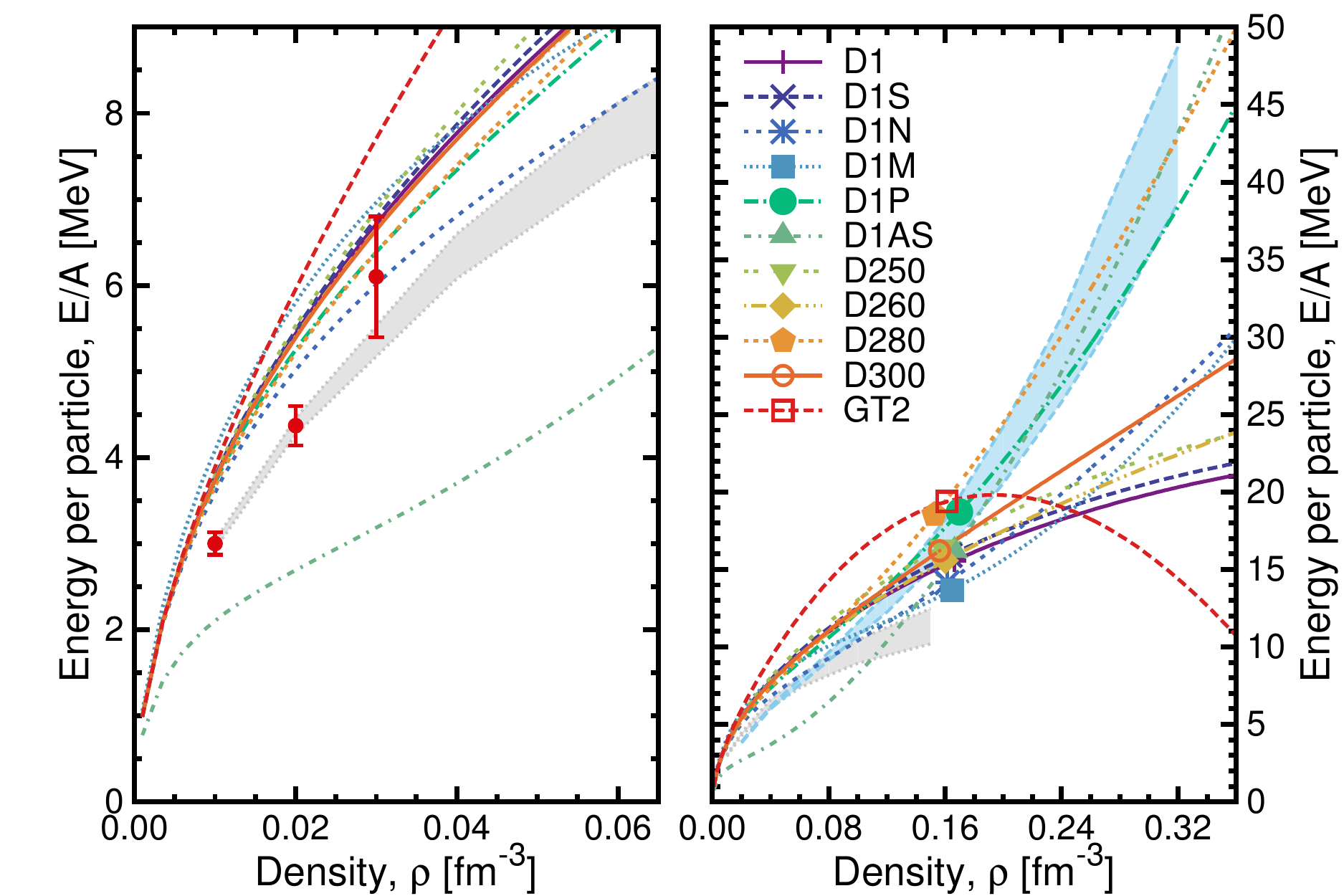}
\caption{(Color online) Energy per particle as a function of density for pure neutron matter for all the Gogny functionals. The shaded region enclosed by a dotted (dashed) line corresponds to quantum Monte Carlo \cite{Gezerlis2013} (self-consistent Green's functions \cite{Carbone2013,*Carbone2014}) calculations based on chiral potentials. The points with error bars on the left panel correspond to three representative points used in Ref.~\cite{Brown2014}. }
\label{fig:ener_pnm}
\end{center}
\end{figure}

As we have seen so far, Gogny functionals are particularly poorly determined in the isovector sector. We therefore start our discussion looking at the energy per particle of neutron matter,
\begin{align}
\label{eq:epnm}
e_\text{PNM}(\rho) =& \frac{3}{5} \frac{\hbar^2k_F^{n,2}}{2 m_n} 
 + \frac{1}{2} \sum_{i=1,2} \Big\{ A^i_1 + C^i_1 \rho^{\alpha_i} \Big\} \rho \nonumber \\
 &+ \frac{1}{2} \sum_{i=1,2}  B^i_{nn} \mathsf{g}\left( \mu_i k_F^n \right) \, ,
\end{align}
where the Fermi momentum is that of neutron matter, $k_F^n=(3 \pi^2 \rho)^{1/3}$. We show in Figure~\ref{fig:ener_pnm} the Gogny functional predictions for the energy per particle of neutron matter. The left panel concentrates on the low-density regime. In addition to the Gogny data, we show results for  microscopic determinations of the energy per particle of neutron matter at low densities. In the left panel, we show a band obtained from a recent Auxiliary Field Diffusion Monte Carlo calculation using N2LO local chiral interactions \cite{Gezerlis2013}. The width of the band reflects the uncertainty in the original interaction, but we note that missing 3 nucleon forces would increase the average value at higher densities. The three points with error bands correspond to the representative data used in Ref.~\cite{Brown2014}. This is obtained as an educated guess, based on a variety of many-body calculations in this low-density regime. 

At sub-saturation densities, we find relatively similar results for the different parametrizations. None of the Gogny functionals  reproduces the two lowest density points of the Brown-Schwenk analysis. A few are able to fit within the relatively large error band of the third. Fitting Skyrme density functionals to nuclei and to these three sub-saturation points in neutron matter yields a good quality density functional in the isovector sector \cite{Brown2014}. This could be a good starting point for future fitting protocols of Gogny functionals. D1AS results go below the many-body data, which indicates that the functional form of Gogny interactions can in principle be amenable to reproduce such low-density points. 

The right panel of Figure~\ref{fig:ener_pnm} concentrates on a wider density regime, up to slightly above twice the saturation density. As expected, the differences between functionals here are quite marked. While around saturation the variations among functionals are accommodated within about $5$ MeV, at $0.24$ fm$^{-3}$ one finds quite larger discrepancies. As a matter of fact, above saturations one can distinguish between two classes of Gogny functionals. On the one hand, a majority of functionals have a rather soft density dependence and predict energies which are well below $30$ MeV for $\rho \lesssim 0.32$ fm$^{-3}$. On the other, D280, D1AS and D1P increase far more quickly with density. This is important, because the density increase will be reflected in a substantially larger pressure which, in turn, will allow for larger and heavier neutron stars. 

Interestingly, the subgroup of Gogny functionals that have a stiff neutron matter energy per particle follows closely the band enclosed by dashed lines. This band is the result of a realistic many-body calculation, obtained within the self-consistent Green's functions approach using an N3LO chiral two-body interaction supplemented with a three-body force at N2LO \cite{Carbone2013,*Carbone2014}. Because of the non-perturbative nature of the resummation scheme, these calculations have been performed with unrenormalized interactions and up to twice the saturation density. Again, the width of the band is a reflection of the uncertainties in the underlying low-energy constants of the chiral interactions. We find that the stiff Gogny functionals are indeed well within this band up to $0.32$ fm$^{-3}$. Note, however, that these very same functionals do not reproduce the many-body results at low densities. 

\begin{figure}
\begin{center}
\includegraphics[width=0.7\linewidth]{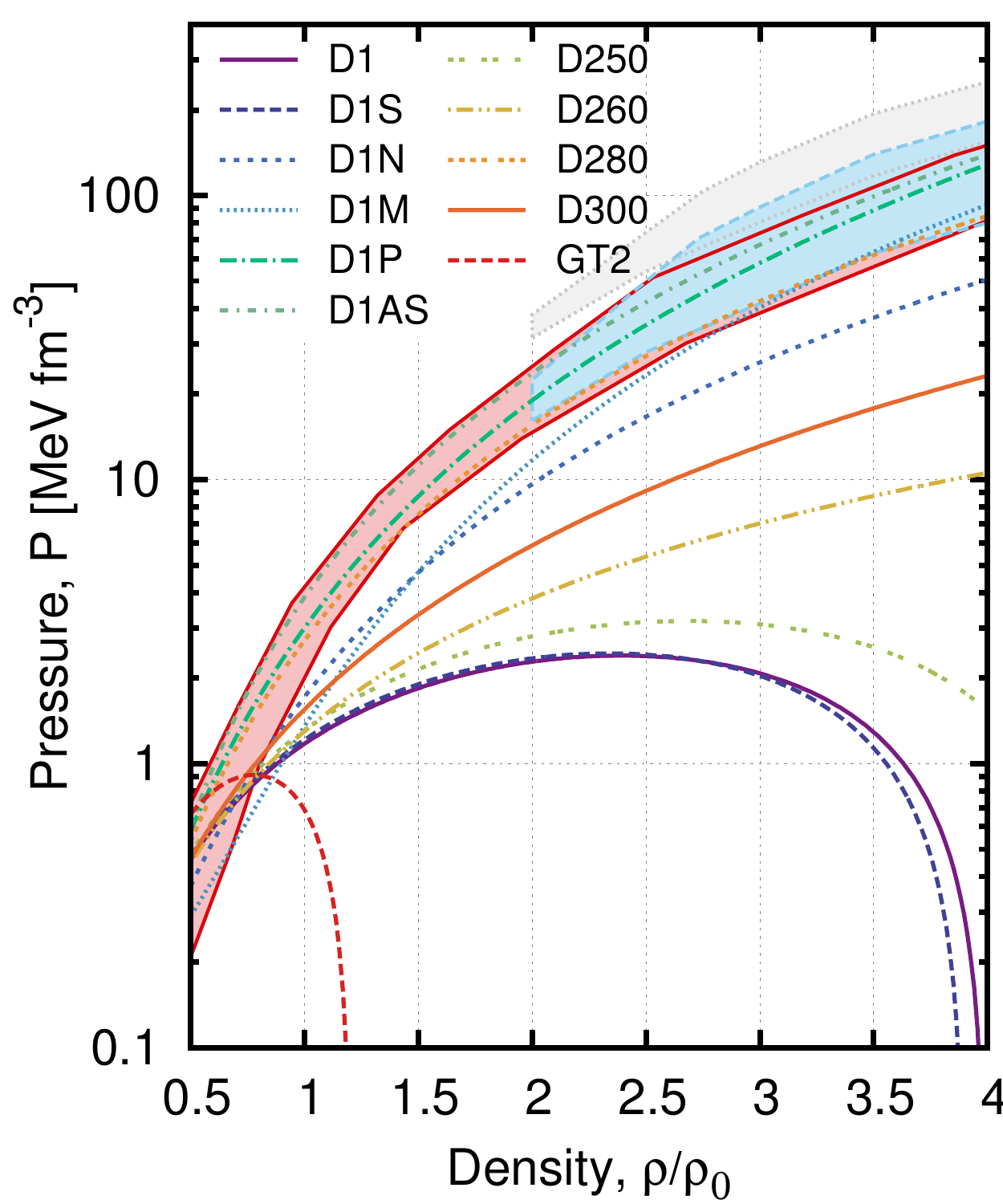}
\caption{(Color online) EoS (pressure as a function of density) for pure neutron matter. The shaded region enclosed by a dotted (dashed) line corresponds to the stiff (soft) symmetry energy obtained with the experimental flow data of Ref.~\cite{Danielewicz2002}. The shaded region enclosed by a full line is obtained from quiescent low-mass X-ray binary mass and radius observations \cite{Lattimer2014b}.}
\label{fig:EoS_PNM}
\end{center}
\end{figure}

GT2 shows a maximum of the energy per particle around saturation, and subsequently decreases as density increases. The decrease of energy as a function of density is a telltale sign of a mechanical instability, which is reflected in a negative pressure. We confirm this fact in Figure~\ref{fig:EoS_PNM}, where we show the pressure in neutron matter for all the Gogny functionals. GT2 indeed shows a negative pressure slightly above saturation. We also find that a few other functionals (D1, D1S, D250) predict negative pressures, although at much larger densities, $\rho \gtrsim 0.64$ fm$^{-3}$ and above. At this point, of course, one should discuss the predictive power of these functionals at such high densities. We do not expect Gogny functionals to describe the physics of the very high density regime, but they might be able to extrapolate some of the low-density nuclear physics into the neutron-star region. In this sense, we use Gogny functionals as effective extrapolation guesses into high densities.  

The pressures of Fig.~\ref{fig:EoS_PNM} are validated agains three sets of constraints. The area enclosed by a solid line is obtained from the data in Fig. 9 of Ref.~\cite{Lattimer2014b}. This delineates the EoS probability distribution 	for neutron stars assuming a  baseline EoS and column densities that have atmospheres containing both  hydrogen and helium, as preferred by some observations \cite{Lattimer2014b}. At densities above twice saturation, we compare the Gogny pressures to the flow data obtained assuming a stiff (grey dotted region) or soft (blue dashed region) symmetry energy \cite{Danielewicz2002}. As expected, the functionals falling close to these constraints are those with the highest values of $L$: D1AS, D1P, and D280. The differences between models in the high-density regime are extreme. Other than the already discussed mechanically unstable models, we find that D260 and D300 have equations of state that are almost an order of magnitude below the empirical constraints. This anomalous behaviour will impact the structure of the corresponding neutron stars, as we shall see next. 

\begin{figure}
\begin{center}
\includegraphics[width=0.7\linewidth]{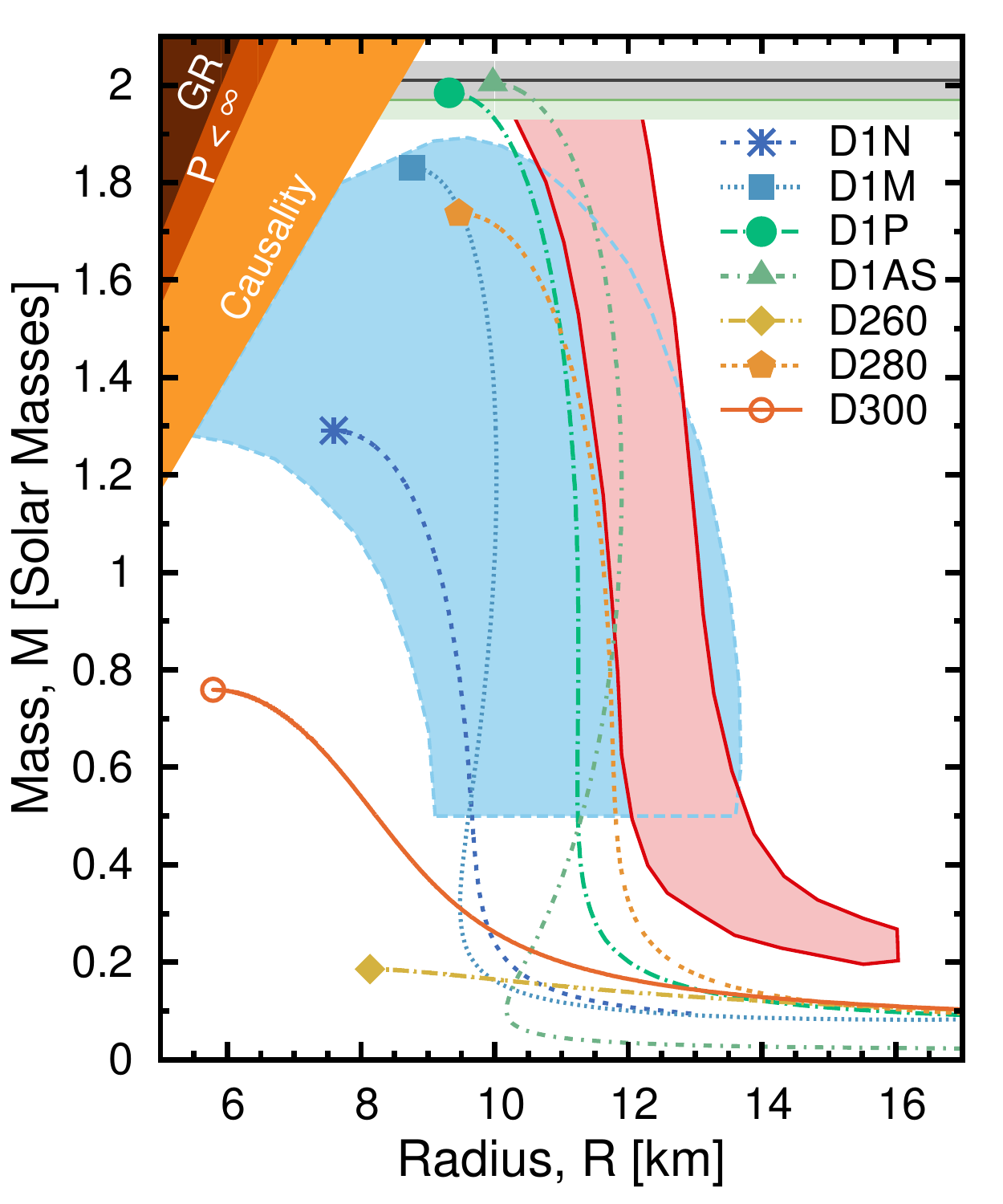}
\caption{(Color online) Mass-radius relation for pure neutron stars obtained with 11 Gogny functionals. The shaded region enclosed by a full line is obtained from quiescent low-mass X-ray binary mass and radius observations using atmosphere models that include both hydrogen and helium \cite{Lattimer2014b}. The shaded region enclosed by a dashed line corresponds to the $90 \, \%$ confidence region for the mass-radius relation of NGC 6397, as obtained in Ref.~\cite{Heinke2014}. }
\label{fig:NS_MR}
\end{center}
\end{figure}

We solved the Tolman--Oppenheimer--Volkov equations to study the structure of pure neutron stars \cite{Shapiro1983}. For simplicity, we considered stars formed only of neutrons.\footnote{We also performed calculations in $\beta-$equilibrium, which yield very similar results.} We stress once again that the results obtained with Gogny functionals at very high densities should be handled with care. We do not take them as realistic estimates, but rather as indications of the strengths and weaknesses of the functionals at the extremes of isospin asymmetry. Also, note that we use the Gogny functional throughout the star, without any low-density matching to a crust EoS. In consequence, our less massive neutron stars have rather small radii. 

We show the resulting mass-radius relations in Figure \ref{fig:NS_MR}. We compare once again with the region (red continuous region) of the most probable mass-radius of Ref.~\cite{Lattimer2014b}. The blue area enclosed by a dashed line corresponds to the $90 \, \%$ confidence region probability contours for the mass and radius of the object NGC 6397 qLMXB, using a helium atmosphere (see right panel of Fig. 9 in Ref.~\cite{Heinke2014}). In addition, we show in Fig.~\ref{fig:NS_MR} the excluded region by causality \cite{Lattimer2012} and the two measurements of neutron-star masses around $M \gtrsim 2 M_\odot$ \cite{Demorest2010,Antoniadis2013}.

Very few functionals predict maximum solar masses close to this limit, and only 4 achieve a maximum mass above $1.4 M_\odot$. Those that do are essentially the parametrizations that fell within the constraints in the energy and the pressure of pure neutron matter.  D1AS has the largest maximum mass, just about $M_\text{max}^\text{D1AS} \approx 2M_\odot$ at $R_\text{max}^\text{D1AS} \approx 10$ km. The maximum mass for D1P raises up to $M_\text{max}^\text{D1P} \approx 1.98 M_\odot$ at $R_\text{max}^\text{D1P} \approx 9$ km. The mass-radius relation for these two functionals is relatively close to the probability distribution of Lattimer \emph{et al.} \cite{Lattimer2014b} and falls right through the constraints of Heinke and coauthors \cite{Heinke2014}. D280, in contrast, falls at the edge of the Lattimer constraints, because its EoS runs just below D1P and D1AS. Above $0.5 M_\odot$, these three forces predict essentially constant radii, between $11$ and $12$ km. At the high mass end, however, D280 falls short and yields a maximum mass of $M_\text{max}^\text{D280} \approx 1.74M_\odot$. 

In addition to these three functionals, D1M also produces a sizeable maximum mass, $M_\text{max}^\text{D1M} \approx 1.83M_\odot$. Since D1M has a substantially softer EoS, though, it reaches this mass by exploring far smaller radii, below $R \approx 10$ km. The only other parametrization that is able to support a neutron star above $1M_\odot$ is D1N. The maximum mass thus obtained is $M_\text{max}^\text{D1M} \approx 1.29M_\odot$. Note, however, that the mass-radius relation is biased towards small radii and falls well below the Lattimer predictions. 

We show two more mass-radius relations in Fig.~\ref{fig:NS_MR}. We already discussed the fact that D300 and D260 have EoS which fall well below the constraints obtained by a variety of empirical analysis. As expected, these two functionals cannot support heavy neutron stars. They also predict rather small objects, with minimum radii of the order of $6-8$ km. Finally, we note that none of the remaining 4 Gogny functionals (D1, D1S, D250, GT2) are able to sustain sizeable neutron stars. One can in principle trust these functionals up to the mechanical instability region, but their maximum masses are all below $0.16 M_\odot$ and their radii lie above $17$ km. These unrealistic neutron star properties are a further indication of poor isovector properties in the Gogny functionals.

\section{Conclusions}

We analysed the Gogny functional predictions for bulk symmetric and isospin asymmetric nuclear matter. We used 11 parametrizations from the literature. This is an extensive set and has allowed for a comprehensive analysis. We provide analytical expressions for all microscopic and macroscopic properties. We note that the exchange term is responsible for non-trivial density and asymmetry dependences. 

There is a different variability of predictions depending on whether one considers the isoscalar or the isovector sector. In general, we find that bulk properties are rather well-constrained in symmetric matter. For single-particle potentials, the isoscalar components are very similar for all functionals in the low-momentum region, below the Fermi surface. To some extent, these results are to be expected, since symmetric matter properties as well as (low-momentum) finite nucleus data are often inputs in the fitting procedure of the functionals. 

In contrast, we find a wide range of results in the isovector channels. At the single-particle level, isovector potentials above and below the Fermi surface can be extremely different, depending on the parametrization under consideration. As a consequence, the effective masses and their isospin asymmetry dependence is somewhat unrestricted. When analysed as a function of momenta, effective masses show a wide range of values. Looking specifically at the isovector splitting of the effective mass at the Fermi surface, we find that most functionals predict $m^*_n > m^*_p$. Having said that, the exact amount and asymmetry dependence is still uncontrolled.

At the bulk level, we analysed the isovector properties predicted by Gogny functionals. In general, we find that the saturation symmetry energy of these functionals falls within empirical estimates. The density dependence, however, is rather soft in most cases, with slope parameters that fall below most empirical estimates. We find that the slope parameter, $L$, is in most cases at its maximum (or decreasing) at saturation. The three functionals with acceptable $L$ parameters (D1P, D1AS, D280) have somewhat extremes values of $S$. This is an artificial tension, and we believe updated fitting protocols could help improve this behaviour. 

An analysis of the different contributions in the isovector channel shows that, by and large, the variability of isovector properties comes from the finite-range matrix elements. For pure neutron matter, we also find a variety of results. At subsaturation, most functionals lie above the region predicted by microscopic many-body calculations. Above saturation, a subgroup of functionals (D1, D1S, D250 and GT2) predicts a mechanical instability. Another subgroup (D260,~D300, D1N, D1M) avoids the instability, but predicts rather low pressures as a function of density. Finally, a third subgroup (D1P, D1AS and D280) have stiffer pressures. As a consequence, the last subgroup can sustain sizeable neutron stars, with radii around $R\approx11-12$ km and maximum masses close to $2M_\odot$. 

These findings point out at a poor fitting strategy, particularly in the isovector sector. The present generation of functionals show a wide variability in the underlying matrix elements of the Gogny interaction, even when they are separated in isospin contributions. Having access to analytical expressions, we identified a few key observables that could be used to improve fitting procedures. The isovector mass splitting at the Fermi surface, for instance, is proportional to $B^i_{nn}$ (and depends mildly on $\mu_i$). The information from optical potential analysis \cite{Charity2014,Li2014} or microscopic calculations \cite{Vidana2009} can be used to constrain this matrix element. Similarly, the high-momentum limit of the isoscalar and isovector single-particle potentials is entirely set by the density-dependent zero-range terms of the functional. By imposing a realistic Lane isovector potential, one could in principle constrain these values. Finally, and possibly more importantly, the bulk isovector properties can nowadays be used to fit the functionals \cite{Erler2013}. By imposing more realistic values of $L$, $K_\tau$ and, if necessary, neutron star properties, the resulting Gogny functionals would provide a better description of isospin-rich systems. 

Because we do not have access to Gogny fitting protocols or data, it is difficult to quantify how much of the variability that we have found  is of a statistical or a systematic nature \cite{Dobaczewski2014}. We suspect, though, that it is the systematic uncertainty that dominates. Most functionals have been fit from the start with rather low symmetry energies, and the corresponding slopes turn out to be unphysically small. Similarly, the density exponents, $\alpha_i$, and the ranges, $\mu_i$, are often guessed initially and not allowed to change. It would be interesting to analyse whether the data prefer other choices of these values. We therefore call for the publication of clearer fitting protocols and the associated correlation matrices. Similarly to what has already been done in Skyrme-like functionals, a covariance  matrix analysis will help quantify, at least, the statistical uncertainty of the data  \cite{Reinhard2010,Kortelainen2010}. In addition, covariance analysis identifies at once which parameters of the fit, if any, are over constrained. 

The results that we presented could be easily extended to study other bulk systems of interests. Having access to the analytical formulas for the single and double Fermi sphere integrals of Eqs.~(\ref{eq:ufun}) and (\ref{eq:h_function}), it is straightforward to extend this analysis to spin-polarised systems. Preliminary studies have analysed the properties of Gogny functionals in extreme spin-polarized neutron matter \cite{Isayev2004,Lopez-Val2006,Perez-Garcia2008}. Fitting these properties to realistic interactions could be a way to constrain the spin sector of the functional. This would be complementary to the traditional analysis based on decomposing the total energy of symmetric matter into spin and isospin channels \cite{Decharge1980,Gogny1977}. 

Another extreme of spin and isospin can be achieved by analysing the case of a single impurity in a Fermi sea of a majority of a different species. In nuclear physics, the case of a spin-up proton embedded in neutron matter has been of recent interest. The quantum many-body problem can be solved in this instance to quite a good degree of accuracy using Monte Carlo techniques \cite{Forbes2014,Roggero2014}. Comparing these results with the analytical expressions predicted by the Gogny functional will provide further means of fitting the spin and the isospin dependence of the effective interaction at the single-particle level. 

We found that some Gogny functionals, both traditional and modern, can lead to isospin or mechanical instabilities - or, in the case of D1, to both. Although this happens at densities beyond those of interest for nuclear physics, the existence and characterisation of these instabilities has only been attempted in a rather generic way \cite{Margueron2002,Margueron2001,Rudra1992}. In particular, the momentum dependence of these instabilities is relevant for Gogny functionals \cite{Margueron2005}. An analysis that goes beyond the low $q$ Landau parameter approximation in isospin asymmetric matter could shed further light into the isovector sector of the Gogny interaction. 

Regarding neutron stars, most Gogny functionals are unable to sustain sufficiently large stars that agree with present astrophysical observations. Mass and radii, however, are just two of the many observational probes of neutron stars. It would be interesting to study further properties, such as cooling, with the few semi-realistic density functionals that sustain sufficiently large neutron stars. In particular, the study of  pairing in isospin asymmetric matter with Gogny forces has not yet been extensively undertaken \cite{Sedrakian2003} and could provide further constraints to the parametrizations. Along similar lines, the extension to finite temperature \cite{Ventura1994,Rios2010a} of some of the present results could provide an insight into isospin transport properties. 

Finally, we note that we have centered our analysis around infinite, bulk matter. As we have seen, this allows for a relatively simple and analytical characterisation of both microscopic and thermodynamical properties. Gogny functionals, however, are specifically designed to reproduce nuclear properties and hence the isovector properties of finite nuclei  should also be probed. Closed shell, isospin-rich nuclei, like $^{48}$Ca, $^{132}$Sn or $^{208}$Pb would be excellent potential starting points for such an analysis \cite{Chappert2008}. Open shell isotopes will explore pairing properties, and the combination of asymmetry and pairing might provide insightful results \cite{Rodriguez2014}. Going further, one might want to explore extensions of the Gogny functional, particularly towards more realistic finite-range density dependent terms \cite{Chappert2007}. 

\appendix

\section{Analytical expressions for single-particle properties}
\label{app:micro}

Single-particle potentials with the Gogny force are obtained by integrating the effective interaction over one Fermi sea in the Hartree--Fock approximation. We express our results in terms of the total density, $\rho$, and the isospin asymmetry parameter, $\beta=\frac{\rho_n - \rho_p}{\rho}$. The Fermi momenta are defined as usual, with:
$ k^n_{F} = k_{F} \left( 1 + \beta \right)^{1/3}$ and 
$ k^p_{F} = k_{F} \left( 1 - \beta \right)^{1/3}$, with 
$k_F = \left( \frac{3 \pi^2}{2} \rho \right)^{1/3}$, the Fermi momentum of the corresponding isospin symmetric system. For both the zero-range and the direct, finite-range terms the single Fermi sphere integration leads to trivial density factors. The integral of the exchange term can be performed analytically and expressed in terms of Gaussian and error functions. One finds the following expression for the single-particle potential \cite{Sedrakian2003,Chen2012}:
\begin{widetext}
\begin{align}
U_\tau(k) = \sum_{i=1,2} \Big\{
\left[ A^i_0 + C^i_0 \rho^{\alpha_i} \right] \rho
+ \tau \left[ A^i_1 + C^i_1 \rho^{\alpha_i+1} \right] \rho \beta 
 +  B^i_{nn} \mathsf{u} \left( \mu_i k, \mu_i k_F^\tau \right) 
 +  B^i_{np} \mathsf{u} \left( \mu_i k, \mu_i k_F^{-\tau} \right)
\Big\} \, .
\label{eq:sp_potential_app}
\end{align}
where we have introduced the dimensionless function:
\begin{align}
\mathsf{u}\left(q ,q_F \right) &= \frac{1}{q} \left[ e^{-\frac{\left(q + q_F \right)^2}{4}} - e^{-\frac{\left(q - q_F \right)^2}{4}} \right] 
+ \frac{\sqrt{\pi}}{2} \left[ \text{erf} \left( \frac{ q + q_F }{2} \right) - \text{erf} \left( \frac{ q - q_F }{2} \right)  \right]  \, .
\label{eq:ufun}
\end{align}
\end{widetext}
The error function is defined as usual,
\begin{align}
 \text{erf} (x) = \frac{2}{\sqrt{\pi}} \int_0^x \text{d} t \, e^{-t^2} \,.
 \label{eq:errorf}
 \end{align}
 The zero-momentum limit of this expression is finite,
\begin{align}
\mathsf{u}\left(0 ,q_F \right) &= - q_F e^{-\frac{q_F^2}{4}} 
+ \sqrt{\pi} \text{erf} \left( \frac{q_F }{2} \right) \, ,
\label{eq:ufun_q0}
\end{align}
and provides information on the single-particle potential at low momentum. Note that the derivative of this expression with respect to $q_F$ is positive, which suggests that the exchange term grows indefinitely with density. The limit $q_F \gg 1$ leads to a constant value, indicating a saturation of the exchange term with density. 

 In some cases of interest, one needs to consider the rearrangement term in the single-particle potential. This is a momentum-independent contribution which arises from the density dependence of the interaction. It is the same for both neutrons and protons and in asymmetric matter it reads:
 \begin{align}
 U_R = \frac{1}{2} \sum_{i=1,2} \left[ C_0^i + C_1^i \beta^2 \right] \rho^{\alpha_i +1} \, .
 \label{eq:rear}
 \end{align}
 Note that the isovector contribution, $C_1^i$, enters with a $\beta^2$ dependence in this term. 
 
With these analytical expressions at hand, one can compute further single-particle properties. The effective mass, for instance, characterises the momentum dependence of the single-particle potential. In asymmetric matter and at arbitrary momentum, it is given by:
 \begin{widetext}
\begin{align}
\frac{m_N(k)}{m_\tau^*} 
= 1 + \frac{m_N}{\hbar^2} \sum_{i=1,2} \mu_i^2 \Big\{ B^i_{nn} \mathsf{m} \left( \mu_i k, \mu_i k_F^{\tau} \right) + 
B^i_{np} \mathsf{m} \left( \mu_i k, \mu_i k_F^{-\tau} \right)  \Big\} \, , 
\label{eq:effective_mass}
\end{align}
 with 
\begin{align}
\mathsf{m} \left( q, q_F \right) = \frac{1}{q^3} \left[  
 \left( 2 - q q_F \right) e^{-\frac{\left(q - q_F \right)^2}{4}} 
-\left( 2 + q q_F \right) e^{-\frac{\left(q + q_F \right)^2}{4}} 
\right] \, .
\label{eq:mfun}
\end{align}
\end{widetext}
This function is essentially obtained as a momentum derivative of Eq.~(\ref{eq:ufun}). The effective mass is often computed at the respective Fermi surface of a given particle species. The term with two equal Fermi momenta reduces to:
\begin{align}
\mathsf{m} \left( q_F, q_F \right) = \frac{1}{q_F^3} \left[  
2 \left( 1 - e^{-q_F^2} \right)   
- q_F^2 \left( 1 + e^{-q_F^2} \right)   
\right] \, .
\end{align}
This function also determines the effective mass at the Fermi surface for symmetric and pure neutron matter. Let us note that, at zero Fermi momentum, both functions, $\mathsf{u}(q,0)=\mathsf{m}(q,0)=0$, vanish. This is relevant for both the low-density and the impurity regimes.

Using the Hugenholtz-van Hove theorem, one can relate the bulk properties of asymmetric nuclear matter to a combination of parameters associated with derivatives of the single-particle potential at the Fermi surface of the symmetric system \cite{Chen2012}. All these parameters can be obtained analytically in the case of the Gogny interaction. In particular, it is useful to separate the isoscalar and isovector contributions of the single-particle potential. The isoscalar potential is obtained in the symmetric matter limit:
\begin{align}
U_0(k) = \sum_{i=1,2} \Big\{
\left[ A^i_0 + C^i_0 \rho^{\alpha_i} \right] \rho
 +  B^i_{0} \mathsf{u} \left( \mu_i k, \mu_i k_F \right) 
\Big\} \, .
\label{eq:isoscalar_sp_potential}
\end{align}
We note that, because both the momentum dependence and the rearrangement terms are non-linear in asymmetry, this isoscalar potential is only approximately equal (but very close) to the combination $\tfrac{U_n(k)+U_p(k)}{2}$ in arbitrarily isospin asymmetric matter.

The isovector potential in nuclear matter can in principle be obtained analogously, from the difference of single-particle contributions. Alternatively, the asymmetry dependence of the single-particle potential can be Taylor expanded. The coefficients of the expansion encode the isospin dependence of the single-particle potentials:
\begin{align}
U_i^\text{sym}(k) &= \frac{1}{i!} \left. \frac{ \partial^i U_n(k)}{\partial \beta^i} \right|_{\beta=0} 
=\frac{(-1)^i}{i!} \left. \frac{ \partial^i U_p(k)}{\partial \beta^i} \right|_{\beta=0} \, .
\label{eq:isovector_sp_potential}
\end{align}
The first coefficient carries most of the information and it is analogous to the Lane potential \cite{Chen2012}. The expression reads:
\begin{align}
U_1^\text{sym}(k) &= 
 \sum_{i=1,2} \Big\{
\left[ A^i_1 + C^i_1 \rho^{\alpha_i} \right] \rho
 + \frac{1}{3} B^i_{1} \mathsf{u_1} \left( \mu_i k, \mu_i k_F \right) 
\Big\} \, ,
\label{eq:u_1_sym}
\end{align}
with
\begin{align}
\mathsf{u_1} \left(q, q_F \right) &= -\frac{q_F^2}{2q} \left[ e^{- \frac{ \left(q + q_F\right)^2}{4} } 
- e^{- \frac{ \left(q - q_F \right)^2}{4}} \right]  \, .
\end{align}
We note that the rearrangement contribution cancels exactly in this case.

%The connection with macroscopic observables is obtained by evaluating these coefficients and some of 
%its momentum derivatives at $k=k_F$. We computed analytically all the coefficients needed to find
%the slope parameter $L$. These read:
%\begin{align}
%\left. \frac{ \partial U_0}{\partial k} \right|_{k_F} k_F &= \sum_{i=1,2} B^i_0 \mathsf{u'_0} \left(\mu_i k_F \right) \\ 
%\left. \frac{ \partial U_1}{\partial k} \right|_{k_F} k_F &= \frac{1}{6} \sum_{i=1,2} B^i_1 \mathsf{u'_1} \left(\mu_i k_F \right) \\ 
%\left. \frac{ \partial^2 U_0}{\partial k^2} \right|_{k_F} k_F^2 &= \frac{1}{2} \sum_{i=1,2}  B^i_0 \mathsf{u'_2} \left(\mu_i k_F \right)
%\end{align}
%with
%\begin{align}
%\mathsf{u'_0} \left(q \right) = \frac{1}{2q} &\left[ 2 - q^2 - \left(2 + q^2  \right) e^{- q ^2 } \right] \, ,\\
%\mathsf{u'_1} \left(q \right) = q &\left[ -1 + \left(1 + q^2  \right) e^{- q ^2 } \right] \, , \\
%\mathsf{u'_2} \left(q \right) = \frac{1}{q} &\left[ -4 + q^2 + \left(4 + 3q^2 +q^4 \right) e^{- q ^2 } \right]  \, .
%\end{align}

\section{Analytical expressions for bulk properties}
\label{app:macro}

At zero temperature, the thermodynamical properties of asymmetric nuclear matter with the Gogny interaction can be expressed analytically. These involve double integrations of the matrix elements over Fermi spheres. Let us start with the energy per particle of asymmetric nuclear matter. The full expression for the Gogny force reads \cite{Chappert2007}:
\begin{widetext}
\begin{align}
e(\rho,\beta) &=
\frac{3}{5}\frac{\hbar^2 k_F^2}{2 m_N} \frac{1}{2} \left\{ (1+\beta)^{5/3} +(1-\beta)^{5/3} \right\}
+ \frac{1}{2} \sum_{i=1,2} \Big\{ \left[ A_i^{0} + C_i^{0} \rho^{\alpha_i} \right] \rho 
+ \left[ A^i_1  + C^i_1 \rho^{\alpha_i} \right] \rho \beta^2 \Big\} \nonumber \\
& + \frac{1}{2} \sum_{i=1,2} \left\{ 
B^i_{nn} 
\left[ \frac{1+\beta}{2} \mathsf{g} \left(\mu_i k_F^n \right) 
     + \frac{1-\beta}{2} \mathsf{g} \left(\mu_i k_F^p \right)  \right] 
+ B^i_{np} 
 \mathsf{h} \left(\mu_i k_F^n,\mu_i k_F^p \right)     
     \right\} \, .
     \label{eq:energy_asymat}
\end{align}
The function $\mathsf{g}(q)$,
\begin{align}
 \mathsf{g} \left( q \right)  = \frac{2}{q^3} - \frac{3}{q} 
 - \left( \frac{2}{q^3} - \frac{1}{q} \right) e^{-q^2} + \sqrt{\pi} \text{erf}  \left( q \right) \, ,
 \label{eq:g_function}
 \end{align}
is the result of a double integration of the exchange matrix elements over the same Fermi surface. It is the sum of a Gaussian and an error function, with the latter given by its standard definition, Eq.~(\ref{eq:errorf}). This function appears as well in the completely degenerate cases and has been quoted in the literature \cite{Decharge1980}. 
The second function,
\begin{align}
 \mathsf{h} \left( q_1, q_2 \right)  = &
 2 \frac{q_1^2 - q_1 q_2 + q_2^2 -2 }{q_1^3 + q_2^3} e^{- \frac{ \left(q_1 + q_2\right)^2}{4} }
- 2 \frac{q_1^2 + q_1 q_2 + q_2^2 -2 }{q_1^3 + q_2^3} e^{- \frac{ \left(q_1 - q_2\right)^2}{4} } \nonumber \\
&- \sqrt{\pi} \frac{q_1^3 - q_2^3}{q_1^3 + q_2^2} \text{erf}  \left( \frac{q_1 - q_2}{2} \right) 
+ \sqrt{\pi} \text{erf}  \left( \frac{q_1 + q_2}{2} \right) \, ,
 \label{eq:h_function}
 \end{align}
 \end{widetext}
is the result of integrating over two different Fermi surfaces and is hence unique to the polarized case. Note that, as expected, $\mathsf{h}$ is a symmetric function of its arguments. In symmetric conditions, $\mathsf{h}$ reduces to $\mathsf{g}$, $\mathsf{h}(q,q)=\mathsf{h}(0,2q) = \mathsf{h}(2q,0) = \mathsf{g}(q)$. 

By using the latter property, one can easily find the energy per particle of the fully (un)polarised cases. In the case of symmetric nuclear matter, one obtains:
\begin{widetext}
\begin{align}
\label{eq:esnm}
e_\text{SNM}(\rho) &= \frac{3}{5} \frac{\hbar^2k_F^2}{2 m_N} 
 + \frac{1}{2} \sum_{i=1,2} \Big\{ A^i_0 + C^i_0 \rho^{\alpha_i} \Big\} \rho 
 + \frac{1}{2} \sum_{i=1,2}  B^i_0 \mathsf{g}\left( \mu_i k_F \right) \, .
\end{align}
Taking derivatives with respect to the density, one can easily find expressions for the pressure, the compressibility and the skewness:
\begin{align}
\label{eq:psnm}
P_\text{SNM}(\rho) &= \rho^2 \frac{\partial e_\text{SNM}}{\partial \rho}= \frac{2}{5} \frac{\hbar^2k_F^2}{2 m_N} \rho 
 + \frac{1}{2} \sum_{i=1,2} \Big\{ A^i_0 + (\alpha_i+1) C^i_0 \rho^{\alpha_i} \Big\}  \rho^2 
 + \sum_{i=1,2} B^i_0 \mathsf{p}\left( \mu_i k_F \right) \rho \, , \\
 \label{eq:k0}
K_{0}(\rho) &= 9 \rho^2 \frac{\partial^2 e_\text{SNM}}{\partial \rho^2} = -\frac{6}{5} \frac{\hbar^2k_F^2}{2 m_N} 
  +  \frac{9}{2} \sum_{i=1,2} (\alpha_i+1) \alpha_i C^i_0 \rho^{\alpha_i+1}
  - 3 \sum_{i=1,2} B^i_0 \mathsf{k}\left( \mu_i k_F \right)  \, , \\
\label{eq:q0}
Q_{0}(\rho) &= 27 \rho^3 \frac{\partial^3 e_\text{SNM}}{\partial \rho^3} = \frac{24}{5} \frac{\hbar^2k_F^2}{2 m_N}
  + \frac{27}{2} \sum_{i=1,2}   (\alpha_i+1) \alpha_i (\alpha_i-1) C^i_0 \rho^{\alpha_i+1} 
  + 3 \sum_{i=1,2} B^i_0 \mathsf{q}\left( \mu_i k_F \right) \, .
\end{align}
We introduced the dimensionless functions:
\begin{align}
\mathsf{p}\left(q \right) &= -\frac{1}{q^3} + \frac{1}{2 q} + \left( \frac{1}{q^3} + \frac{1}{2 q} \right) e^{- q^2} \, , \\
\mathsf{k}\left(q \right) &= -\frac{6}{q^3} + \frac{2}{q} + \left( \frac{6}{q^3} + \frac{4}{q} + q\right) e^{- q^2} \, , \\
\mathsf{q}\left(q \right) &= -\frac{54}{q^3} + 14 q + \left( \frac{54}{q^3} + \frac{40}{q} + 13 q + 2 q^2 \right) e^{- q^2} \, ,
\end{align}
\end{widetext}
which are a combination of derivatives of the original $\mathsf{g}$ function over its arguments. We note here that neither $K_0$ nor $Q_0$ receive any contribution from the direct term of the finite range part.

In the case of isospin-imbalanced matter, Eq.~(\ref{eq:energy_asymat}) provides the full expression for the energy. Successive derivatives of this function over density and isospin asymmetry yield the isovector bulk properties of the system. These can, again, be written analytically. The expressions read:
\begin{widetext}
\begin{align}
\label{eq:srho}
S(\rho) &= \frac{1}{3} \frac{\hbar^2k_F^2}{2 m_N} 
 +\frac{1}{2} \sum_{i=1,2} \Big\{ A^i_1 + C^i_1 \rho^{\alpha_i} \Big\} \rho 
 + \frac{1}{6} \sum_{i=1,2} \Big\{ B^i_{nn}  \mathsf{s}_1\left( \mu_i k_F \right)
 +  B^i_{np}  \mathsf{s}_2\left( \mu_i k_F \right) \Big\} \, , \\
 \label{eq:lrho}
L(\rho) &= \frac{2}{3} \frac{\hbar^2k_F^2}{2 m_N}
 + \frac{3}{2} \sum_{i=1,2} \Big\{A^i_1 + (\alpha_i+1) C^i_1 \rho^{\alpha_i} \Big\} \rho
 + \frac{1}{6} \sum_{i=1,2} \Big\{  B^i_{nn} \mathsf{l}_1\left( \mu_i k_F \right)
 +   B^i_{np}   \mathsf{l}_2\left( \mu_i k_F \right) \Big\} \, , \\
  \label{eq:ksymrho}
 K_\text{sym}(\rho) &= -\frac{2}{3} \frac{\hbar^2k_F^2}{2 m_N} 
          +\frac{9}{2}  \sum_{i=1,2} (\alpha_i+1) \alpha_i C^i_1 \rho^{\alpha_i+1} 
 - \frac{2}{3} \sum_{i=1,2}  \Big\{ B^i_{nn} \mathsf{k}_1\left( \mu_i k_F \right)
 + B^i_{np} \mathsf{k}_2\left( \mu_i k_F \right) \Big\} \, ,
\\
  \label{eq:qsymrho}
Q_\text{sym}(\rho) &= \frac{8}{3} \frac{\hbar^2k_F^2}{2 m_N}
   + \frac{27}{2} \sum_{i=1,2} (\alpha_i+1) \alpha_i (\alpha_i-1) C^i_1 \rho^{\alpha_i+1}
 + \frac{1}{3} \sum_{i=1,2} \Big\{ B^i_{nn} \mathsf{q}_1\left( \mu_i k_F \right)
 +  B^i_{np}  \mathsf{q}_2\left( \mu_i k_F \right) \Big\} \, ,
\end{align}
with the functions:
\begin{align}
\label{eq:s12}
\mathsf{s}_1\left(q \right) &= \frac{1}{q} - \left( \frac{1}{q} + q \right) e^{- q^2} \, , \quad
\mathsf{s}_2\left(q \right) = \frac{1}{q} - q - \frac{1}{q} e^{- q^2} \, , 
\\
\label{eq:l12}
\mathsf{l}_1\left(q \right) &= - \frac{1}{q} + \left( \frac{1}{q} + q + 2q^3 \right) e^{- q^2} \, , \quad
\mathsf{l}_2\left(q \right) = - \frac{1}{q} - q + \left( \frac{1}{q} + 2q \right) e^{- q^2} \, , 
\\
\mathsf{k}_1\left(q \right) &= - \frac{1}{q} + \left( \frac{1}{q} + q + \frac{q^3}{2} + q^5 \right) e^{- q^2} \, , \quad
\mathsf{k}_2\left(q \right) = - \frac{1}{q} - \frac{q}{2} + \left( \frac{1}{q} + \frac{3}{2} q + q^3\right) e^{- q^2} \, , 
\\
\mathsf{q}_1\left(q \right) &= - \frac{14}{q} + \left( \frac{14}{q} + 14 q + 7 q^3 + 4 q^5 + 4q^7 \right) e^{- q^2} \, , \quad
\mathsf{q}_2\left(q \right) = - \frac{14}{q} - 5 q + \left( \frac{14}{q} + 19 q + 12 q^3 + 4q^5 \right) e^{- q^2} \, .
\end{align}
\end{widetext}
We note that $\mathsf{s}_i$ and $\mathsf{l}_i$ ($\mathsf{k}_i$ and $\mathsf{q}_i$) are proportional to $q^3$ ($q^5$) in the limit $q \to 0$. Moreover, the limits are such that the exchange terms are determined entirely by $B_1^i = B^i_{nn}  - B^i_{np}$. In the opposite extreme, $q \gg 1$, the linear term in $\mathsf{s}_2$, $\mathsf{l}_2$, $\mathsf{k}_2$ and $\mathsf{q}_2$ dominates. In particular, this implies that in the high density limit: (a) the exchange contributions grow indefinitely as $\rho^{1/3}$ and (b) the matrix elements $B^i_{np}$ determine the high-density behaviour of the exchange term of isovector properties.

\acknowledgments

This work is supported by STFC through Grants ST/I005528/1, ST/J000051/1 and ST/L005816/1. Partial support comes from "NewCompStar", COST Action MP1304.

\bibliographystyle{apsrev4-1}
\bibliography{biblio}

\end{document}